\documentclass[a4paper,11pt]{article}
\pdfoutput=1
\usepackage{jheppub}
\usepackage[T1]{fontenc}

\usepackage{mathtools}
\usepackage{mathrsfs}
\usepackage[capitalise]{cleveref}
\usepackage[dvipsnames]{xcolor}
\usepackage{siunitx}
\DeclareSIUnit{\year}{yr}
\DeclareSIUnit{\au}{AU}
\DeclareSIUnit{\parsec}{pc}
\usepackage[normalem]{ulem}
\usepackage{subcaption}
\usepackage[version=4]{mhchem}
\usepackage{slashed}
\newcommand{\refcite}[1]{Ref.~\cite{#1}}
\newcommand{\refscite}[1]{Refs.~\cite{#1}}
\usepackage{aas_macros}
\usepackage{booktabs}

\usepackage{bm}
\newcommand{\bb}[1]{\bm{\mathrm{#1}}}
\newcommand{\du}{\mathrm{d}}
\newcommand{\dd}{\,\du}
\newcommand{\sm}{\mathrm{SM}}
\newcommand{\dm}{\mathrm{DM}}
\newcommand{\ddm}{\textnormal{DDM}}
\newcommand{\dark}{\mathrm{D}}
\newcommand{\disk}{\mathrm{disk}}
\newcommand{\vir}{\mathrm{vir}}
\newcommand{\gas}{\mathrm{gas}}
\newcommand{\halo}{\mathrm{halo}}
\newcommand{\core}{\mathrm{core}}
\newcommand{\cloud}{\mathrm{cloud}}
\newcommand{\ion}{\mathrm{ion}}
\newcommand{\rec}{\mathrm{rec}}
\newcommand{\exchange}{\mathrm{ex}}
\newcommand{\total}{\mathrm{tot}}
\newcommand{\molecular}{\mathrm{mol}}
\newcommand{\atomic}{\mathrm{atom}}
\newcommand{\milkyway}{\mathrm{MW}}
\newcommand{\sech}{\operatorname{sech}}
\newcommand{\circular}{\mathrm{circ}}
\newcommand{\jeans}{\mathrm{J}}
\newcommand{\Gaia}{\textit{Gaia}}
\newcommand{\dco}{\textnormal{DCO}}
\newcommand{\pr}{\mathrm{p}}
\newcommand{\peri}{\mathrm{p}}
\newcommand{\el}{\mathrm{e}}
\newcommand{\dpr}{\mathrm{X}}
\newcommand{\del}{\mathrm{c}}
\newcommand{\be}{E_{\mathrm{B}}}
\newcommand{\chandrasekhar}{\mathrm{Ch}}
\newcommand{\cool}{\mathrm{cool}}
\newcommand{\PI}{\mathrm{PI}}
\newcommand{\eq}{\mathrm{eq}}
\newcommand{\height}{Z}
\newcommand{\sound}{\mathrm{s}}
\newcommand{\pair}[2]{\{#1#2\}}
\newcommand{\binary}{\mathrm{bin}}
\newcommand{\sampled}{\mathrm{MC}}

\begin{document}

\title{\boldmath \Gaia{}'s dark partners:\\A new signature of dissipative dark matter}

\author[a,b]{Badal Bhalla,}
\emailAdd{badal.bhalla@utah.edu}
\affiliation[a]{Department of Physics and Astronomy, University of Utah, Salt Lake City, UT 84112, USA}
\affiliation[b]{Homer L. Dodge Department of Physics and Astronomy, University of Oklahoma, Norman, OK 73019, USA}

\author[c,d]{David Dunsky,}
\emailAdd{ddunsky@umd.edu}
\affiliation[c]{Center for Cosmology and Particle Physics, Department of Physics, New York University, New York, NY 10003, USA}
\affiliation[d]{Maryland Center for Fundamental Physics, Department of Physics,
University of Maryland, College Park, MD 20742, USA}

\author[e]{Akshay Ghalsasi,}
\emailAdd{aghalsasi@fas.harvard.edu}
\affiliation[e]{Jefferson Physical Laboratory, Department of Physics, Harvard University, Cambridge, MA 02138, USA}

\author[f,g,1]{Benjamin V. Lehmann,\note{Einstein Fellow}}
\emailAdd{benvlehmann@gmail.com}
\affiliation[f]{Kavli Institute for Astrophysics and Space Research, Massachusetts Institute of Technology, Cambridge, MA 02139, USA}
\affiliation[g]{Center for Theoretical Physics---a Leinweber Institute, Massachusetts Institute of Technology, Cambridge, MA 02139, USA}

\author[e]{Jackie Lodman,}
\emailAdd{jackie.lodman@gmail.com}

\author[b]{and Kuver Sinha}
\emailAdd{kuver.sinha@ou.edu}

\preprint{\ignorespaces
    \parbox[t]{4.5cm}{\raggedleft
        MIT-CTP/6091}}
\date\today


\abstract{\ignorespaces
The \Gaia{} mission has greatly improved our understanding of the stellar binary population. Beyond providing an extensive catalog of binary systems with two luminous stars, \Gaia{} has also discovered several systems that appear to be star--black hole (BH) binaries. One of the leading candidate formation channels for these systems is dynamical exchange, in which one component of a star--star binary is ejected from the system and replaced by a BH. Exchange processes are very sensitive to the velocity distribution of the incoming objects, and become very efficient in the limit that perturbers are slow, so the exchanged binary population is a sensitive probe of slow-moving objects. A population of slow-moving, nonluminous objects is naturally produced in dissipative dark matter (\ddm{}) scenarios, where a subcomponent of dark matter can cool to form a dark disk and dark compact objects (\dco{}s), including but not limited to BHs, with a small velocity dispersion. In this work, we evaluate the rate at which \dco{}s formed by \ddm{} exchange into star--star binaries to form star--\dco{} binaries. We find that in a wide range of parameter space, even under conservative assumptions, the rate of exchange interactions is sufficient to produce the objects that \Gaia{} has observed to date. In some portions of parameter space, the predicted exchange rate is well in excess of the rate estimated from observations. We point out the potential for competitive constraints on the \ddm{} parameter space with current and future data, making the \Gaia{} mission an important testbed for dark sector cooling.
}

\maketitle
\flushbottom

\section{Introduction}
Despite decades of observational and experimental efforts, there is still no direct evidence for any nongravitational interactions between dark matter (DM) and Standard Model (SM) species~\cite{Bertone:2004pz,Profumo:2017hqp,Cirelli:2024ssz}. Many ongoing searches continue to target well-motivated nongravitational interactions, but the community must now acknowledge a troubling possibility: it may be that the DM resides in a fully secluded sector, with any interactions suppressed to such a degree that they will remain inaccessible to any realistic experiment. Accordingly, it is now essential to develop new methods of probing DM physics leveraging gravity alone.

This case is not as hopeless as it might seem. Everything that is currently known about DM has been inferred strictly from gravitational observables. Moreover, there are many possible deviations from the benchmark $\Lambda$--cold dark matter ($\Lambda$CDM) model that might be communicated to the SM by gravity alone. The simplest class of modifications are those in which the DM is subject to nongravitational interactions within the dark sector. DM self-interactions famously modify the behavior of DM halos, with consequences for the formation and behavior of observable baryonic structures~\cite{Tulin:2017ara}. Thus, even if no interaction bridges DM and the SM, such self-interactions are gravitationally detectable. On yet smaller scales, gravitational observables constrain DM in the form of massive compact objects such as primordial black holes (PBHs)~\cite{Carr:2009jm,Carr:2020gox,Carr:2016drx,Carr:2020xqk,Carr:2021bzv,Carr:2026hot}. In fact, a dynamical upper bound on the mass of the DM constituent is arguably the first limit to have ever been placed on DM~\cite{1969Natur.224..891V}.

There is, however, another class of interactions that might lead to observable features: namely, those that lead to dissipation. The importance of dissipative interactions is demonstrated prominently by the SM itself. Baryons form structures such as stars and galaxies that are dramatically different from CDM structures, and this difference is entirely due to dissipation. CDM halos reach virial equilibrium gravitationally, and maintain a high virial temperature, with no path to cool. The SM gas cools radiatively, losing energy and collapsing into disks, clumps, and compact objects.
It is not possible for the entirety of the DM to cool in this way, but in a complex dark sector, it is entirely possible that a fraction of the DM does cool and collapse. This would lead to the formation of dark structures, such as disks, that would be very different from CDM halos, and potentially detectable by their gravitational effects within galaxies~\cite{Fan:2013tia,Fan:2013yva,Fan:2013bea,Buch:2018qdr,Ghalsasi:2017jna,Roy:2023zar,Gemmell:2023trd,Kramer:2016dqu,Kramer:2016dew,Schutz:2017tfp,Widmark:2018ylf,Widmark:2020vqi,Widmark:2021gqx}.

But the galactic scale is not the only setting in which DM cooling has significant effects. The same dissipative dark matter (\ddm{}) models that produce structures such as dark disks also enable cooling on a \emph{stellar} scale. As \ddm{} cools, it behaves similarly to baryonic gas, fragmenting at a mass scale of order the Jeans mass $M_\jeans$. These fragments cool further to produce compact, massive clumps, and possibly even black holes (BHs). Regardless of the nature of these collapsed objects, they behave dynamically as point masses. We will refer to them in this work as \textit{dark compact objects} (\dco{}s).

This presents a tantalizing opportunity: as a host of tools are already being developed to probe the local population of BHs, they might also be applied to probe dissipative dark sector dynamics. This has contributed to the motivation for gravitational wave searches for subsolar-mass BHs: any BH lighter than the Chandrasekhar mass~\cite{Chandrasekhar:1931ih} cannot originate from stellar evolution, and would therefore signal physics beyond the SM, \textit{e.g.}\ PBHs~\cite{Carr:2016drx,Carr:2020xqk,Carr:2021bzv,Carr:2026hot} or BHs formed from dissipative dark sectors~\cite{Shandera:2018xkn,Chang:2018bgx,Gurian:2022nbx,Fernandez:2022zmc}. More generally, a wide variety of PBH probes are incidentally sensitive to \dco{}s from dark sectors with efficient cooling. Na\"ively, only a small portion of the \ddm{} parameter space can be constrained in this way: at most a few percent of the DM can be strongly dissipative, so PBH constraints are only useful where they are already strong enough to probe small PBH DM fractions.

There is, however, an important gap in this argument: the \emph{phase space distribution} of generic PBHs is quite different from that of \dco{}s produced by \ddm{}\@. The kinematics of PBHs are essentially indistinguishable from CDM. But \ddm{} objects inherit the distribution of the \ddm{} gas in which they are born, and this distribution is strongly modified by efficient cooling. This means that when considered in the Galactic context, these \ddm{} objects tend to be slow-moving and overdense towards the Galactic center. Thus, any observable which is enhanced for slow-moving objects in the vicinity of baryonic stars will be more sensitive to \ddm{} objects than to PBHs of the same mass.

There is one class of \dco{} probes that behaves in exactly this way: scattering processes with stellar binaries. In particular, consider dynamical exchange, in which a \dco{} interacts with a binary system and ejects and replaces one of the two components, forming a new binary system with one star and one \dco{}\@. Dynamical exchange is known to already be marginally sensitive to PBH DM~\cite{Bhalla:2024jbu}, in that exchange processes between PBHs and components of stellar binaries can account for the observed abundance of star--\dco{} binaries at the order-of-magnitude level. The exchange process also has a cross section $\Sigma_{\exchange}$ that increases sharply at low relative velocity, scaling as $\Sigma_{\exchange} \sim v^{-2}$. This combination suggests that \emph{observations of star--\dco{} binaries} could provide a novel and powerful probe of \ddm{} objects.

Fortuitously, star--\dco{} binaries have recently become observationally accessible thanks to the \Gaia{} mission~\cite{Gaia:2016zol,El-Badry:2024vjt}. The \Gaia{} observatory is a space telescope designed to precisely measure proper motions and radial velocities of a large number of stars in the Milky Way (MW). These results have already provided invaluable information for modeling evolution and properties of the MW itself~\cite{Cautun:2019eaf,Posti:2019tbp,Eilers:2019gqs}, and have enabled refined measurements of the potential of the MW halo. \Gaia{} has also greatly expanded the sample of known stellar binaries within the MW, and has further identified at least three star--\dco{} binary systems, generally presumed to be star--BH binaries, and thus labeled \Gaia{} BH1--3~\cite{El-Badry:2022zih,Chakrabarti:2022eyq,El-Badry:2023pah,Gaia:2024ggk}.

These systems, and particularly BH1 and BH3, have generated considerable interest in the astrophysics literature, since their properties would be nontrivial to produce via isolated stellar evolution~\cite{Rastello:2023fkx,DiCarlo:2023qey,Tanikawa:2023svu,Kotko:2024oaw}. Several authors have thus proposed that these systems might originate from exchange between a stellar binary and an exotic compact object, on the premise that the formation rate due to ordinary stellar evolution might eventually be deemed insufficient. (However, see \refcite{Iorio:2024pat} for a contrary argument.) In particular, the exchange rate between stars and PBH DM is large enough, at the order-of-magnitude level, to account for BH1 and BH3. Given that the exchange rate between stars and PBHs is already large enough to be potentially observable when the PBHs have a velocity distribution typical of a CDM halo, this process represents a perfect opportunity to probe BHs formed by \ddm{}: here, even as the number density of these objects is reduced, their lower velocity dispersion should significantly enhance the exchange rate. If this enhancement is large enough, \ddm{} could account for BH1 and BH3, or could even be in tension with the low measured abundance of star--BH binaries.

The abundance and properties of such systems will soon be probed much more comprehensively by the upcoming fourth data release from \Gaia{} (\Gaia{} DR4). Accordingly, in this work, we evaluate the rate of star--BH binary formation in \ddm{} scenarios, with appropriate modifications to the number density and phase space distribution. We determine portions of parameter space in which \Gaia{} is plausibly sensitive to the \dco{} population from \ddm{}, and identify regions in which \ddm{} could account for the star--\dco{} binaries observed to date.

This work is organized as follows. In \cref{sec:dissipative-dark-matter}, we introduce our benchmark \ddm{} model, and relate microphysical parameters to cooling rates and the resultant dark disk formation and compact object masses. In \cref{sec:three-body}, we describe our calculation of exchange rates at low \dco{} velocity, and we validate these estimates with numerical simulations. We present our results in \cref{sec:results}, and we discuss the implications in \cref{sec:discussion}.

\section{Dissipative dark matter and compact objects}
\label{sec:dissipative-dark-matter}
In order to understand the population of dark compact objects exchanged into binary systems, we must connect the properties of those objects to the underlying dissipative DM model. Exchange is ultimately governed by three distributions of compact object properties whose basic features we must predict: (1) their spatial distribution, (2) their momentum distribution, and (3) their mass distribution. Each of these three distributions is nontrivially related to the microphysical properties of the DM\@, and we study each one in turn in the following three subsections.

In broad terms, the behavior of the dissipative DM fluid is determined by the rate and mechanism of cooling as a function of environmental properties. If cooling is efficient, the \ddm{} gas will cool to form a rotationally supported gaseous dark galactic disk. If the rotationally supported gaseous disk is unstable to fragmentation, the gas can clump and collapse to form dark compact objects in the form of \ddm{} stars,\footnote{While these objects are analogous to SM stars, we will use ``dark compact objects'' throughout the remainder of this work to allow for the possibility that these objects collapse to black holes, and to avoid confusion with ``dark stars'' in the sense of \refcite{Spolyar:2009nt}, which are \textit{baryonic} stars partially powered by DM annihilation.} analogous to SM stars. (We assume there are no nuclear processes in the dark sector, so these ``stars'' do not undergo fusion at their cores.) The formation and fragmentation of the disk, and thus all of the distributions of interest, are ultimately controlled by the cooling rate of the \ddm{} gas.
In principle, the dark sector can accommodate many different cooling channels, along with nuclear feedback, which could produce a complex cooling history and dark stellar mass function.
In order to make concrete predictions, we choose a well-known benchmark model for the remainder of this work, bearing in mind that other classes of dissipative models might generate similar objects.

\subsection{DDM and disk formation}
\label{sec:benchmark-model}
We consider the DM model known in the literature as ``atomic DM''~\cite{Kaplan:2009de}. Here, a DM subcomponent has particle content and interactions similar to SM baryons, \textit{i.e.}, a dark proton with mass $m_{\dpr}$ and a dark electron with mass $m_{\del}$, charged under dark electromagnetism, with a dark photon coupling with strength $\alpha_\dark$. We assume a Lagrangian of the form
\begin{equation}
    \mathcal L \supset
    -\tfrac14 {F_{\dark}}_{\mu\nu} F_{\dark}^{\mu\nu}
    + \bar{\dpr}\left(
        i\slashed{D} - m_\dpr
    \right) \dpr
    + \bar\del\left(
        i\slashed{D} - m_\del
    \right) \del
    ,
\end{equation}
where $F_\dark^{\mu\nu}$ is the dark field-strength tensor and $D_\mu$ is the gauge-covariant derivative for a dark $\operatorname{U}(1)$ symmetry under which $\dpr$ has charge $+1$ and $\del$ has charge $-1$.
In particular, this allows the dark sector to form dark atoms and molecules as bound states of dark protons and dark electrons. We assume that the dissipative dark matter (\ddm{}) makes up a fraction of all DM given by $\epsilon_\ddm \equiv \Omega_\ddm/\Omega_\dm$.

\ddm{} has been explored extensively in the literature~\cite{Goldberg:1986nk,Kaplan:2009de, Kaplan:2011yj, Cyr-Racine:2012tfp, Cline:2012is,Cline:2013pca, Fan:2013yva,Fan:2013tia,Fan:2013bea,Cyr-Racine:2013fsa,Foot:2014osa,Randall:2014kta,Randall:2014lxa,Foot:2016wvj,Rosenberg:2017qia, Ghalsasi:2017jna,Chang:2018bgx,Shandera:2018xkn,Gresham:2018anj,Essig:2018pzq,Alvarez:2019nwt,Roux:2020wkp,Cyr-Racine:2021oal,Cline:2021itd,Chacko:2021vin,Ryan:2021dis,Gurian:2021qhk,Ryan:2021tgw,Howe:2021neq,Blinov:2021mdk,Bansal:2021dfh,Cruz:2022otv,Peled:2022byr,Roy:2023zar, Gemmell:2023trd,Kouvaris:2015rea, Giudice:2016zpa, Curtin:2019lhm, Hippert:2021fch, Gross:2021qgx, Ryan:2022hku, Gurian:2022nbx, Armstrong:2023cis}. Here we review the elements that are essential to the formation of \dco{}s. When the \ddm{} gas falls into a DM halo, it is shock-heated to the virial temperature given by \refcite{Loeb:2006za} 
\begin{equation}
    \label{eq:Tvir}
    T_\vir
    \simeq \qty{32}{\electronvolt}
    \left(\frac{M_\halo}{\qty{e10}{M_\odot}}\right)^{2/3}
    \left(\frac{m_{\dpr}}{\qty{1}{\giga\electronvolt}}\right)
    \left(\frac{1+z}{10}\right)
    \, .
\end{equation}
Dark atoms then cool through atomic transitions, bremsstrahlung, and inverse Compton scattering with dark photons in the dark equivalent of the cosmic microwave background\@. In this work, we will remain agnostic about the background dark photon temperature, and thus neglect inverse-Compton cooling. The cooling rates for each process are given by \refcite{Rosenberg:2017qia}. When the temperature is well above the atomic binding energy $\be$, these cooling timescales are given approximately by
\begin{align}
    \label{eq:tbrem}
    &t_{\mathrm{brem}} \simeq
    \frac{9\left(\frac{3\pi}{2}\right)^{1/2}m^{3/2}_{\del}T_\vir^{1/2}}{32\alpha_\dark^3 n_\dpr x_\ion^{2}}
    \approx \qty{9e6}{\year} \left(\frac{m_{\del}}{m_{\el}}\right)^{3/2}
    \left(\frac{T_\vir}{\qty{35}{\electronvolt}}\right)^{1/2}
    \left(\frac{\alpha_\sm}{\alpha_\dark}\right)^{3}
    \left(\frac{\qty{1}{\centi\meter^{-3}}}{n_\dpr x_\ion^2}\right)
    ,
    \\&
    \label{eq:tatomic}
    t_{\atomic} \simeq
    \frac{9 \left(\frac{3\pi}{2}\right)^{1/2}T_\vir^{2}}
        {32\alpha^{6}_{\dark}n_\dpr (1-x_\ion)}
    \approx \qty{e7}{\year}
    \left(\frac{T_\vir}{\qty{35}{\electronvolt}} \right)^2
    \left(\frac{\alpha_\sm}{\alpha_\dark} \right)^{6}
    \left(\frac{\qty{1}{\centi\meter^{-3}}}{n_\dpr (1-x_\ion)} \right),
\end{align}
where $n_\dpr$ denotes the dark proton number density and $x_\ion$ denotes their ionization fraction.
In practice, we use the full forms of the atomic cooling rates from \refcite{Rosenberg:2017qia} rather than the approximate forms above.

In order for the \ddm{} gas to cool efficiently, the temperature needs to be at least comparable to $\be/10$. This implies that the \ddm{} gas cools efficiently only for $T \gtrsim \alpha_\dark^2 m_{\del}/20$. Prior to such a phase of efficient cooling, the gas tracks the virial temperature of the halo (\cref{eq:Tvir}), which evolves on the dynamical timescale $t_{\mathrm{dyn}}$. The protons shock-heat, and communicate their temperature to electrons via Rutherford scattering, on a timescale $t_\eq$. The dynamical and equilibration timescales are given by~\cite{Fan:2013yva, Outmezguine:2018nce}
\begin{equation}
    \label{eq:tdyn}
    t_{\mathrm{dyn}} = \frac{1}{\sqrt{8\pi G \rho}}
    \,,\qquad
    t_\eq = \frac{m_{\dpr} m_{\del}}{2\sqrt{3}\alpha_\dark^2n_\dpr}
    \left(\frac{3T_{\vir}}{m_{\del}}\right)^{3/2}
    \left[\log\left(
        1+\frac{T^{2}_{\vir}}{4\alpha_\dark^{2}n_\dpr^{2/3}}\right)\right]^{-1}
    \,,
\end{equation}
where we have assumed $T = T_{\vir}$ in the above expression. In order to consider cooling to be efficient, we will require
\begin{equation}
    \label{eq:coolcond}
    t_\eq < t_\cool
    \equiv
    \min(t_{\mathrm{brem}},t_{\atomic})  < t_{\mathrm{dyn}}
    .
\end{equation}
\par
Since the virial temperature of the halo is controlled by the halo mass as a function of redshift, the evolutionary history of the halo is central to the cooling of the \ddm{} and the growth of the dark galaxy. As with SM baryons, the \ddm{} will form a dark galaxy with disk, bulge, and halo components. Systematic investigations of the dark galaxy morphology have been performed using merger tree simulations~\cite{Ghalsasi:2017jna} and, for select points in the parameter space, full hydrodynamical simulations~\cite{Roy:2023zar}. In this work, we simply consider the evolution of the average main progenitor of a MW-like halo with $M_\halo = \qty{e12}{M_\odot}$ at $z=0$. We use this final mass to calculate the growth history of the average main progenitor using the formalism developed in \refcite{vandenBosch:2014wng}, and thus infer the virial temperature of the halo as a function of time. While this procedure does not correctly capture the full merger history of the MW halo, it is sufficient for present purposes, and we leave a merger tree simulation for future work.

\begin{figure}
    \centering
    \includegraphics[width=0.8\linewidth]{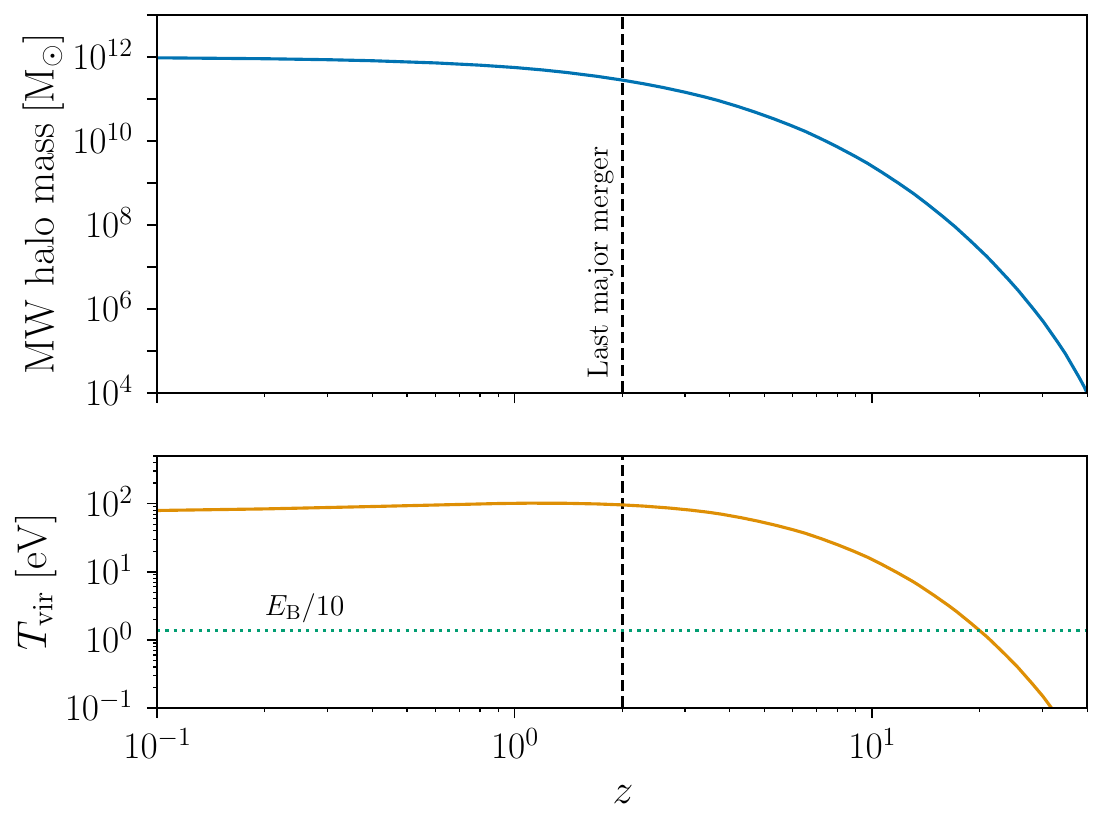}
    \caption{The evolution of the average main progenitor of the Milky Way halo with mass \qty{e12}{M_\odot} at $z = 0$. We also include the mass of the \ddm{} disk after the last major merger (any disk structure formed previously would have been disrupted), assuming 1\% of the DM is \ddm{} in the dark thin disk ($\epsilon_\disk=0.01$). Here $\be$ is set to the binding energy of SM hydrogen.}
    \label{fig:progenitor-evolution-atomic}
\end{figure}

The evolution of the main progenitor mass and virial temperature is shown in \cref{fig:progenitor-evolution-atomic}. As the progenitor grows and the virial temperature increases, $T_\vir$ eventually becomes appreciable relative to $\be$, crossing the threshold of $\alpha^2_{\dark} m_{\del}/20$ to ionize the gas and produce efficient cooling. As the virial temperature increases further, cooling becomes progressively inefficient, as the number density of the dark atoms in a virialized halo decreases as ${\sim} (1+z)^3$. Cooling is efficient only for a particular window of cosmic time. During this period, the cooled dark atoms form a galactic disk. Major mergers can destroy the galactic disk and form a spheroidal bulge. The MW is thought to have experienced its last major merger at $z\simeq 2$~\cite{2018MNRAS.478..611B,2018Natur.563...85H,2020ApJ...901...48N}, so we only consider the Galactic disk formed from the gas which would cool at $z < 2$.

As the halo grows through mergers, the baryons and \ddm{} that are added to the halo cool and contribute to the growth of the dark galaxy (see \cref{fig:progenitor-evolution-atomic}). We compute the dark disk mass $M_{\disk}^{\ddm}$, radius $R_{\disk}^{\ddm}$, and height $\height_{\disk}^{\ddm}$ at $z = 0$ by integrating the amount of \ddm{} gas that cools after the last major merger of the MW. Specifically, for each parameter point $(\alpha_\dark, m_{\del}, m_{\dpr}, \epsilon)$, we use the following procedure: beginning from the last major merger at $z=2$, we take a step $\delta z$, corresponding to a time $\delta t$. In that time, the fraction of the \ddm{} gas which cools is na\"ively given by $\delta t/t_\cool(z)$. However, only the gas which equilibrates is available for cooling, a fraction $\delta t/t_\eq(z)$. Thus, the fraction of the gas which actually cools in this step is given by
\begin{equation}
    \delta f_\cool(z, \delta z) =
        \min\left\{1,\;\frac{\delta t(z, \delta z)}{t_\eq(z)}\right\}
        \times\min\left\{1,\;\frac{\delta t(z, \delta z)}{t_\cool(z)}\right\}
    .
\end{equation}
This cooled \ddm{} gas will contribute in an amount $\delta M_\disk^\ddm = \delta f_\cool(z, \delta z) \times \epsilon_{\ddm} \times 0.2\times M_\halo(z)$. The factor of $0.2$ accounts for the fraction of \ddm{} that is found in the dark thin disk rather than in the thick disk or bulge of the dark galaxy~\cite{Roy:2023zar}.

We further assume that this contribution has the phase space distribution of a disk formed at redshift $z$. Such a disk has radius $R_\disk^\ddm(z) = 2^{-1/2} \lambda R_\vir(z)$~\cite{Mo:1997vb}, where $\lambda$ is a dimensionless spin parameter with a log-normal distribution with a mean value of $0.039$, and $R_\vir$ is the virial radius of the halo, given by 
\begin{equation}
    \label{eq:rvir}
    R_\vir(z) = \qty{33.6}{\kilo\parsec} \times
    \left(\frac{M_\halo(z)}{\qty{e10}{M_{\odot}}}\right)^{1/3}
    \left(\frac{2}{1+z}\right)
    .
\end{equation}
The height of the gaseous disk at redshift $z$ is given by~\cite{2008gady.book.....B,Krumholz:2005mt}
\begin{equation}
    \height_\disk^\ddm(z) =
    \frac{\sigma_{\height}^2 \left(R_{\disk}^{\ddm}(z)\right)^2}
        {G\left[
        M_{\disk}^{\ddm} + M_{\disk}^{\sm}
        \right]
        }
    ,
\end{equation}
where $\sigma_\height$ is the vertical velocity dispersion. For the initial gaseous disk that forms, the velocity dispersion $\sigma_\height$ is controlled by the temperature of the gas after cooling, \textit{i.e.}, $\sigma_{\height,T} = \sqrt{3T/m_{\dpr}}$. Once the dark gaseous disk becomes Toomre-unstable and clumps to form \dco{}s, gravitational interactions give an additional contribution $\sigma_{\height,\mathrm{grav}}$ to the velocity dispersion.

\begin{figure}
    \centering
    \includegraphics[width=\linewidth]{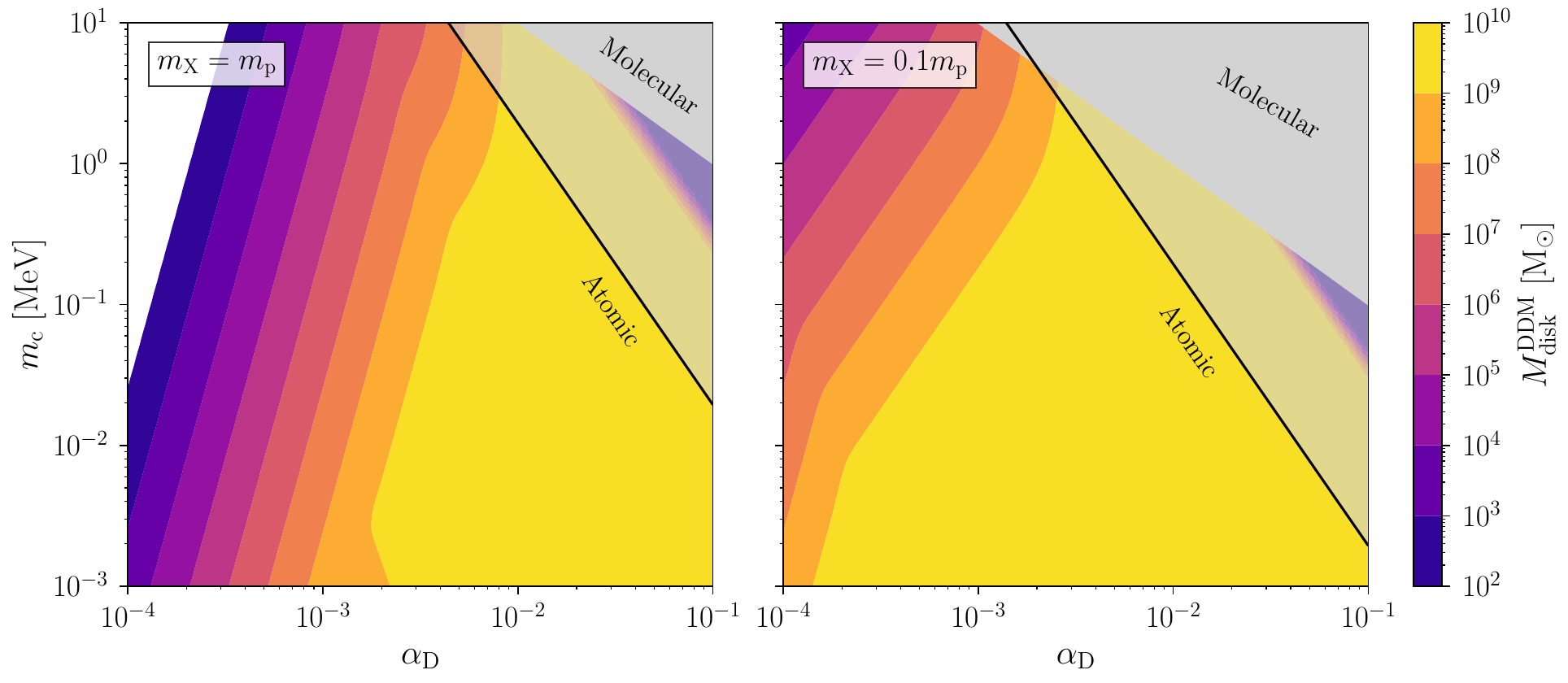}
    \caption{The mass of the \ddm{} disk for our benchmark parameters $\epsilon_\ddm=0.05$, $\epsilon_\disk=0.01$, and $m_\dpr = m_\pr$. The left and right panels show disk masses assuming dominance of molecular and atomic cooling, respectively. In the gray regions, the height of the \ddm{} disk exceeds that of the MW disk assuming atomic cooling (black line and light shaded region) or molecular cooling (gray shaded region), and our thin disk approximation breaks down.}
    \label{fig:ddm-disk-mass}
\end{figure}

The procedure above defines a differential dark disk with mass $\delta M_\disk^\ddm(z, \delta z)$, radius $R_\disk^\ddm(z)$, and height $\height_\disk^\ddm(z)$. Each disk has a density profile of the form~\cite{Buch:2018qdr, Fan:2013yva}
\begin{equation}
    \delta\rho_\disk^\ddm(z, \delta z;\,R, \height) =
    \frac{\delta M_{\disk}^{\ddm}(z, \delta z)\;e^{-R/R_{\disk}^{\ddm}(z)}}
        {8\pi \bigl(R_\disk^\ddm(z)\bigr)^2 \height_{\disk}^{\ddm}(z)}
    \;\sech^2\left(\frac{\height}{2\height_{\disk}^{\ddm}(z)}\right)
    .
\end{equation}
In practice, as we will explain shortly, $\height_\disk^\ddm = \height_\disk^\sm$ at all points where we compute exchange rates. We compute the parameters of the dark disk at $z=0$ by integrating over such differential disks in $n$ steps. The density profile at $z=0$ is then given by
\begin{equation}
    \label{eq:density}
    \rho_{\disk}^{\ddm}(R, \height) =
    \sum_{n=0}^{2/\delta z}
    \delta\rho_\disk^\ddm(z_n, \delta z;\,R, \height)
    ,
\end{equation}
where $z_n = 2 - n\,\delta z$. The final disk mass as a function of $\alpha_{\dark}$ and $m_{\del}$ is shown for $m_\dpr =  m_\pr$ and $m_\dpr = 0.1 m_\pr$ in \cref{fig:ddm-disk-mass}.

The late-time disk radius $R_\disk^\ddm$ and disk height $\height_\disk^\ddm$ are each computed from $R_{1/2}$ and $\height_{1/2}$, the radius and height containing half of the mass of the disk when the other coordinate is unbounded. We set $R_\disk^\ddm = R_{1/2}$ directly. We find that the \ddm{} disk radius is approximately the same as the SM disk radius, and is only weakly dependent on the \ddm{} parameters. Thus we choose the dark disk radius today to be the same as that of the SM disk, \textit{i.e.}, $R^\ddm_\disk = R^\sm_ \disk = \qty{2.15}{\kilo\parsec}$ (see \cref{tab:scan-parameters}).\footnote{This value is the mass-weighted scale length, which differs slightly from the half-mass radius that we use in our procedure.} The late-time height of the gaseous disk is initially set by the temperature the gas can cool down to. However, once the gaseous disk becomes Toomre-unstable and fragments into dark compact objects (DCOs), gravitational interactions will give rise to a further velocity dispersion. The final velocity dispersion is then given by
\begin{equation}
    \sigma_{\height}^2 =
    \sigma_{\height, \mathrm{grav}}^2+\sigma^2_{\height, T},
\end{equation}
where
\begin{equation}
    \label{eq:sigmazgrav}
    \sigma_{\height, \mathrm{grav}} \approx
    \frac{\sqrt{G M_\disk \height_{\disk}}}{ R_{\disk}}
\end{equation}
is the velocity dispersion coming from gravitational interactions~\cite{2021MNRAS.506.1761S}. Throughout our parameter space of interest, the dark disk mass is subdominant to the SM disk for our fiducial $\epsilon_\ddm = 0.05$.
We assume that the gravitational coupling between the SM disk and the \ddm{} disk is sufficient to set $\sigma_{\height,\mathrm{grav}}^{\ddm} = \sigma_{\height, \mathrm{grav}}^{\sm}$.
This results in conservative estimates for exchange rates, since in most of the parameter space, $\sigma_{\height}^\ddm$ will otherwise be smaller than $\sigma_{\height}^\sm$, enhancing the exchange rate. Note, however, that sufficiently thin dark disks are subject to observational constraints~\cite{Widmark:2021gqx}, so if this assumption is violated and the dark disk remains much thinner than the SM disk, some of the parameter space we study may be excluded.
The gray regions of \cref{fig:ddm-disk-mass} show the region in which $\sigma_{\height,T}$ exceeds $\sigma_{\height,\mathrm{grav}}$. (This is equivalent to the region in which $\height_\disk^\ddm > \height_\disk^\sm$.) The solid gray region assumes that the dark disk temperature is set by molecular cooling to $T = \alpha_\dark^2 m_\del^2/m_\dpr$, and the light shaded region corresponds to atomic cooling. In that region, the dark disk height becomes large, and the thin disk approximation breaks down. Equivalently, cooling of the \ddm{} gas becomes inefficient in this region, and we do not expect many DCOs.

We are interested in parameter space that forms a large dark disk in order to enhance the rate of exchanges with binaries. This requirement translates to a set of restrictions on the parameter values, and ultimately forces the \ddm{} parameters to be comparable to the SM parameters. Specifically, we restrict ourselves to the following parameter space:
\begin{itemize}
    \item \textit{Dark fine structure constant:} $\num{e-4} \leq \alpha_\dark \leq \num{e-1}$. For smaller values, cooling is too inefficient to form a large disk. For larger values, the spectrum of the dark hydrogen atom is modified by relativistic corrections, and our calculation must be modified.
    \item \textit{Dark electron mass:} $\qty{e-3}{\mega\electronvolt} \leq m_{\del} \leq \num{e-1} m_{\dpr}$. For smaller values, the dark electrons would become relativistic at the virial temperature of the halo. For larger values, the dark electron mass becomes comparable to the dark proton mass, violating an assumption of many of the arguments above---although note that since we neglect nuclear physics, the mass is the only distinction between $\del$ and $\dpr$, so a scenario with $m_\del \gg m_\dpr$ behaves identically to one with $m_\del \ll m_\dpr$.
    \item \textit{Dark proton mass:} $0.1m_\pr \leq m_{\dpr} \leq \qty{10}{\giga\electronvolt}$. At lower values, the virial temperature falls below $\be$, so dark atomic hydrogen is never ionized. At higher values, $t_\eq$ becomes large, violating the condition in \cref{eq:coolcond}.
\end{itemize}
We adopt fiducial values $\epsilon_\ddm = 0.05$ and $0.1$ for estimates in the remainder of this work. Comparing dynamical masses of the MW galaxy with mass measured using stellar luminosity does not allow a dark galaxy larger than the measured baryonic mass of the MW galaxy, so $\epsilon_\ddm = 0.05$ is marginally safe with respect to this constraint. (See \textit{e.g.}\ Fig.~4 of \refcite{Roy:2023zar} for the particular parameter space points chosen there.) A value of $\epsilon_\ddm$ much larger than 0.05 will be subject to constraints on dark galaxy mass or to dark disk constraints~\cite{Buch:2018qdr}.

\subsection{Dynamics of the dark thin disk}
\label{sec:thin-disk}
Even if $\epsilon_\ddm=0.05$, not all of the \ddm{} will be found in the dark thin disk as was assumed in \refcite{Fan:2013yva}. Simulations in \refcite{Roy:2023zar} found that only about 20\% of the \ddm{} could be found in the dark thin disk, with the remainder found in the dark thick disk and bulge. We denote the fraction of DM that is in the form of \ddm{} in the dark thin disk as $\epsilon_\disk$.\footnote{For notational simplicity, the subscript ``disk'' refers only to the thin disk throughout this work. In particular, $\epsilon_\disk$ does not include DM in the thick disk.} Therefore, $\epsilon_\ddm = 0.05$ corresponds to $\epsilon_\disk\approx0.2\times\epsilon_\ddm\approx0.01$. As the hydrodynamical simulations of \refcite{Roy:2023zar} were performed for only two points in \ddm{} parameter space, this may be a conservative choice.

We find that the disk profile at $z=0$ is well-approximated by a single thin-disk profile of the form of \cref{eq:density} (\textit{i.e.}, the form given by \refscite{Buch:2018qdr, Fan:2013yva}) with disk parameters $R_{\disk}^{\ddm}$ and $\height_{\disk}^{\ddm}$ determined by the procedure in \cref{sec:benchmark-model}. We thus take
\begin{equation}
    \label{eq:rhoDDM}
    \rho_\disk^\ddm(R, \height) =
    \frac{\epsilon_\disk M_\halo^\milkyway}
        {8\pi \left(R_{\disk}^{\ddm}\right)^2 \height_{\disk}^{\ddm}}
    \exp\left(-\frac{R}{R_{\disk}^{\ddm}}\right)
    \sech^2\left(\frac{\height}{2\height_{\disk}^{\ddm}}\right)  
    .
\end{equation}
The Jeans equations for an isothermal sheet then give the velocity distribution for such a disk in the form
\begin{equation}
    \label{eq:velocity-distribution-cylindrical}
    \mathcal{F}_\text{sheet}^{(i)}(v_{R,i}, v_{\phi,i}, v_{\height,i}) =
    \frac{1}{2\pi^{3/2}\sigma_{R,i}^2\sigma_{\height,i}}
    \exp\left(
        -\frac{v_{R,i}^2}{2\sigma_{R,i}^2}
        -\frac{\tilde{v}_{\phi,i}^2}{\sigma_{R,i}^2}
        -\frac{v_{\height,i}^2}{2\sigma_{\height,i}^2}
    \right)
    \,,
\end{equation}
where $i\in\{\sm, \ddm\}$, and coordinates are cylindrical with $\tilde{v}_{\phi,i} \equiv v_\circular-v_{\phi,i}$~\cite{2008gady.book.....B}, with $v_\circular$ denoting the circular velocity at the position of the Solar System. In the remainder of this work, we denote the position of the Solar System by coordinates $(R, \height) = (R_\odot, \height_\odot)$. This velocity distribution is a multivariate Gaussian with central values $(0, v_\circular, 0)$ and standard deviations $(\sigma_R, \sigma_R/\sqrt{2}, \sigma_{\height})$ in $(R, \phi, \height)$ respectively. The velocity dispersion $\sigma_{\height,i}$ for both the SM and \ddm{} disks is set by gravitational interactions and is given by \cref{eq:sigmazgrav}. We then set $\sigma_R$ using the relation $\sigma_R^2=\sigma_{\height}^2/0.3$, as observed in the SM disk~\cite{2021MNRAS.506.1761S}. Parameter values for the SM disk are given in \cref{tab:scan-parameters}.

\subsection{Dark compact objects}
\label{sec:ddmstars}
The formation of dark compact objects (\dco{}s), such as black holes and dark analogs of stars, is a complicated process. The formation and evolution of the first stars in the Standard Model (Pop.\ III stars) is an area of active research (see \refscite{Bromm:2003vv,Bromm:2013iya,Klessen:2023qmc} for reviews). However, this case has been studied for decades by methods ranging from analytical estimates~\cite{Tegmark:1996yt,Tan:2003bs,McKee:2007yx} to sophisticated hydrodynamical simulations~\cite{Bromm:2001bi,Abel:2001pr,Yoshida:2008gn,Greif:2011it,Hosokawa:2011qa,Hirano:2013lba,Susa:2014moa,Hirano:2015wxa,2016MNRAS.462.1307S}, and the basic mechanism of Pop.\ III star formation is now well understood. This mechanism can be mapped onto the \ddm{} case, enabling us to estimate the properties of the resulting \dco{} masses. The detailed calculation will be presented by a subset of the authors of the present work in \refcite{future-mass-function}. Here we outline the steps and calculate the mass of \ddm{} stars assuming that molecular cooling is efficient.

In order for dark compact objects, \dco{}s, such as \ddm{} stars or BHs to form, gravity must overcome any thermal and rotational support in the dark gas. A rotating gaseous disk is gravitationally unstable if the Toomre criterion is satisfied, \textit{i.e.},
\begin{equation}
    \label{eq:Q}
    Q \equiv \frac{c_\sound \kappa}{\pi G \Sigma} < 1
    ,
\end{equation}
where $Q$ is the Toomre parameter; $\kappa \simeq \sqrt{2}\Omega$ is the epicyclic frequency of the disk with angular velocity $\Omega$; $\Sigma$ is the surface density of the gaseous disk; and $c_\sound \approx \sqrt{T_{\gas}/m_{\dpr}}$ is the sound speed of the gas at temperature $T_{\gas}$. Since we consider a disk that consists of two components, one of SM and one of \ddm{}, the Toomre criterion is modified, and becomes easier to satisfy~\cite{1984ApJ...276..114J}. Accordingly, we find that if gas is able to cool and collapse into the disk in the first place (\textit{i.e.}, $T_{\gas} \lesssim T_{\vir}$), then the Toomre criterion is satisfied across our parameter space of interest. 

The gravitational instability leads to the formation of dark clumps, akin to giant molecular clouds seen in galaxies. The evolution of the collapse and the properties of the resulting stars are determined by the gas temperature $T$ and the dark hydrogen number density $n_{\dpr}$. In particular, the size of collapsing fragments is set by the competition between self-gravity and pressure, and can be crudely estimated by the Jeans radius and mass at the temperature of the halo, given respectively by
\begin{equation}
    R_\jeans = \frac{1}{2}\left(
        \frac{15 T}{4 \pi G m_\dpr^{2} n_\dpr}\right)^{1/2}
    ,\qquad
    M_\jeans = \frac{4\pi}{3} m_{\dpr} n_{\dpr} R^{3}_{\jeans}
        \propto  \frac{T^{3/2}}{n_{\dpr}^{1/2}}
    \;.
\end{equation}
The Jeans mass, $M_\jeans$, then sets the maximum size the star can achieve. The Jeans mass of the clumps that initially fragment due to atomic cooling is given by~\cite{Krumholz:2005mt, future-mass-function}
\begin{multline}
    \label{eq:initial-clump-mass}
    M_\cloud^\atomic
    \approx \frac{\pi^3 \sigma^3 Q}{6 G \Omega}
    \\
    \approx  \qty{6e6}{M_\odot} \left(
        \frac{T_\gas}{\frac{1}{20}\alpha_\dark^2 m_\del}
    \right)^{3/2}
    \left(\frac{\alpha_\dark}{\alpha_\sm}\right)^3
    \left(\frac{m_\del}{m_\el}\right)^{3/2}
    \left(\frac{m_\dpr}{m_\pr}\right)^{-3/2}
    \left(\frac{H|_{z=2}}{H(z)}\right) Q
    \, ,
\end{multline}
where $H$ denotes the Hubble parameter.
In the absence of molecular cooling, \textit{i.e.}, if the cooling is purely atomic, the gas cools adiabatically, the free-fall velocity always remains subsonic, and further fragmentation does not happen~\cite{Guszejnov:2018hdj}, as seen in simulations of atomically-cooled halos~\cite{Bromm:2002hb, Latif:2013dua}. Thus, for atomic cooling, $M_\jeans = M_\cloud^\atomic$.

Further cooling of the gas will cause the clumps to fragment to smaller masses. If molecular cooling is efficient, then the gas rapidly cools further and the speed of sound drops rapidly, allowing for further fragmentation. The Jeans mass is then set by the lowest temperature and density that the gas reaches before cooling becomes inefficient. The lowest temperature to which molecules can cool is set by rotational transitions in molecular hydrogen and can be written as~\cite{Ghalsasi:2017jna}
\begin{equation}
    \label{eq:Tmol}
    T_{\mathrm{rot}} = \qty{300}{\kelvin} \times
        \left(\frac{\alpha_{\dark}}{\alpha_\sm}\right)^{2}
        \left(\frac{m_{\del}}{m_{\el}}\right)^{2}
        \left(\frac{m_{\pr}}{m_{\dpr}}\right)
    .
\end{equation}
\par
The temperature of the gas will only be set by $T_{\mathrm{rot}}$ if \ce{\dpr{}2} is sufficiently abundant. During halo virialization, the gas heats up and ionizes, and \ce{\dpr{}2} molecules unavoidably form if the gas is not exposed to any strong sources of radiation that can dissociate them~\cite{Oh:2001ex}. The abundance of \ce{\dpr{}2} is controlled by a balance of several processes~\cite{1961Obs....81..240M,Bromm:2001bi}. Molecules are produced through radiative association, $\ce{\dpr{}-} + \ce{\dpr{}} \rightarrow \ce{\dpr{}2} + \ce{\del{}-}$, which in turn requires production of \ce{\dpr{}-}. This takes place via radiative attachment, $\ce{\dpr{}} + \ce{\del{}-} \rightarrow \ce{\dpr{}-} + \ce{\gamma}$. However, radiative attachment requires \ce{\del{}-}, and free \ce{\del{}-} are also consumed by recombination, $\ce{\dpr{}+} + \ce{\del{}-} \rightarrow  \ce{\dpr{}} + \ce{\gamma}$. The equilibrium abundance of \ce{\dpr{}2} is thus set by the ratio of the rates of radiative attachment and recombination, leading to a frozen-out molecular fraction of~\cite{Bromm:2000nn, Bromm:2001bi, Tegmark:1996yt, Oh:2001ex}
\begin{equation}
    \label{eq:fX2Abundance}
    f_{\ce{\dpr{}2}} \approx
        \frac{\langle \sigma v\rangle_{\mathrm{attachment}}}
            {\langle \sigma v\rangle_{\mathrm{recombination}}}
    \sim \num{e-3} \,.
\end{equation}
This fraction is independent of \ddm{} parameters~\cite{future-mass-function}, and we therefore treat it as fixed in what follows. Molecular formation and cooling in \ddm{} has been studied in \refscite{Ryan:2021dis,Gurian:2021qhk,Ryan:2021tgw}. However, the formation of molecules including feedback is beyond the scope of this work. Thus we consider two distinct cases, where molecular cooling is efficient and where only atomic cooling is efficient.

As the density increases, the gas fragments onto smaller  and smaller scales until cooling becomes inefficient once molecules reach local thermal equilibrium. This is known as the ``loitering phase''~\cite{Bromm:2001bi}. In the SM, the loitering phase occurs at a number density of $n_{\ce{H},\mathrm{L}} \approx \qty{e4}{\centi\meter^{-3}}$, and in the \ddm{} case, we have~\cite{Ghalsasi:2017jna,Bromm:2001bi}
\begin{equation}
    \label{eq:nHloitering}
    n_{\dpr{},\mathrm{L}} \approx
    \qty{e4}{\centi\meter^{-3}} \times
    \left(\frac{\alpha_{\dark}}{\alpha_\sm}\right)^{8}
    \left(\frac{m_{\del}}{m_{\el}}\right)^{7}
    \left(\frac{m_{\pr}}{m_{\dpr}}\right)^{4}
    .
\end{equation}
The final mass of the gas cloud is then set by the Jeans mass in the loitering phase, and is given by~\cite{future-mass-function,Ghalsasi:2017jna}
\begin{equation}
    \label{eq:Mmax}
    M_\cloud^\molecular \approx \qty{500}{M_{\odot}} \times
    \left(\frac{\alpha_\sm}{\alpha_{\dark}}\right)
    \left(\frac{m_{\el}}{m_{\del}}\right)^{1/2}
    \left(\frac{m_{\pr}}{m_{\dpr}}\right)^{3/2}
    .
\end{equation}
Thus the maximum mass a star can have is given by
\begin{equation}
     \label{eq:maxDDMStar}
    M_\cloud =
    \begin{dcases}
        \qty{500}{M_\odot}
        \left(\frac{\alpha_\dark}{\alpha_\sm}\right)^{-1}
        \left(\frac{m_\del}{m_\el}\right)^{-1/2}
        \left(\frac{m_\dpr}{m_\pr}\right)^{-3/2}
            & T_{\cloud}^{\min} = T_{\mathrm{rot}} \, ,
        \\
        \qty{6e6}{M_\odot} \left(\frac{\alpha_\dark}{\alpha_\sm}\right)^{3}
        \left(\frac{m_\del}{m_\el}\right)^{3/2}
        \left(\frac{m_\dpr}{m_\pr}\right)^{-3/2}
        \left(\frac{H|_{z=2}}{H(z)}\right)
            &  T_{\cloud}^{\min} = \be/10 \, ,
    \end{dcases}
\end{equation}
where we have set $Q \approx 1$ in the second line.
While the above gives the \textit{maximum} mass a star can have, the star does not always reach this mass due to feedback which can shut off accretion, or from preemptively forming a BH by the pair-instability mechanism. Below, we explain the stellar evolution and quantify the maximum mass of the star. 

Since cooling is efficient, the fragmented cloud relaxes into a nearly isothermal profile with a large central density. Eventually, however, cooling becomes opacity-limited: when the central density is sufficiently large, the mean free path of emitted photons is too short for radiation to escape, and the dense core cools only via emission from its surface, rather than throughout its volume. This stalls the growth of the central density, forming a small protostar at the center of the cloud with initial mass given by~\cite{10.1093/mnras/176.3.483}
\begin{equation}
    \label{eq:Mproto}
    m_\dco^0 \simeq \num{e-2} M_{\chandrasekhar}
    \times
    \left(\frac{\alpha_{\dark}}{\alpha_\sm}\right)^{1/2}
    \left(\frac{m_{\del}}{m_{\el}}\right)^{1/4}
    \left(\frac{m_{\pr}}{m_{\dpr}}\right)^{1/4}
    \left(\frac{T}{\be{}}\right)^{1/4}
    ,
\end{equation}
where $M_\chandrasekhar = \qty{5.6}{M_\odot} \times (m_{\pr}/m_{\dpr})^{2}$ is the Chandrasekhar mass. The surrounding gas in the cloud can then accrete onto the protostar. However, since the gas does not efficiently lose angular momentum, it forms an accretion disk around the protostar~\cite{Tan:2003bs,McKee:2007yx,Hosokawa:2011qa}. The protostar--accretion disk system grows through spherical Bondi-like accretion at a rate given by\ignorespaces
\footnote{The protostar--disk system grows through spherical accretion. However, the protostar itself accounts for an $\mathcal{O}(1)$ fraction of the mass of the system, so we model the star itself as growing through Bondi accretion. The resulting growth formula is a simplification. In our implementation, we follow \refcite{Tan:2003bs} and model the growth of the star with the correct density profile for the gas, which deviates slightly from an isothermal profile.}
\begin{equation}
    \label{eq:Maccretion}
    \dot m_\dco \approx  \frac{c_{\sound}^3}{G}
    \simeq \frac{1}{G}
    \left(\frac{T_\gas}{m_{\dpr}}\right)^{\tfrac32} = 
    \begin{dcases}
    \qty{4e-4}{\frac{M_{\odot}}{\year}}
        \left(\frac{\alpha_{\dark}}{\alpha_{\sm}}\right)^3
        \left(\frac{m_{\del}}{m_{\el}}\right)^3
        \left(\frac{m_{\dpr}}{m_{\pr}}\right)^{-3}
            \quad \hfill \textnormal{(molecular cooling)}\\
       \qty{1}{\frac{M_{\odot}}{\year}}
        \left(\frac{\alpha_{\dark}}{\alpha_{\sm}}\right)^3
        \left(\frac{m_{\del}}{m_{\el}}\right)^{\tfrac32}
        \left(\frac{m_{\dpr}}{m_{\pr}}\right)^{-\tfrac32}
            \quad \hfill \textnormal{(atomic cooling).}
    \end{dcases}
\end{equation}
\par
\begin{figure}
    \centering
    \includegraphics[width=\linewidth]{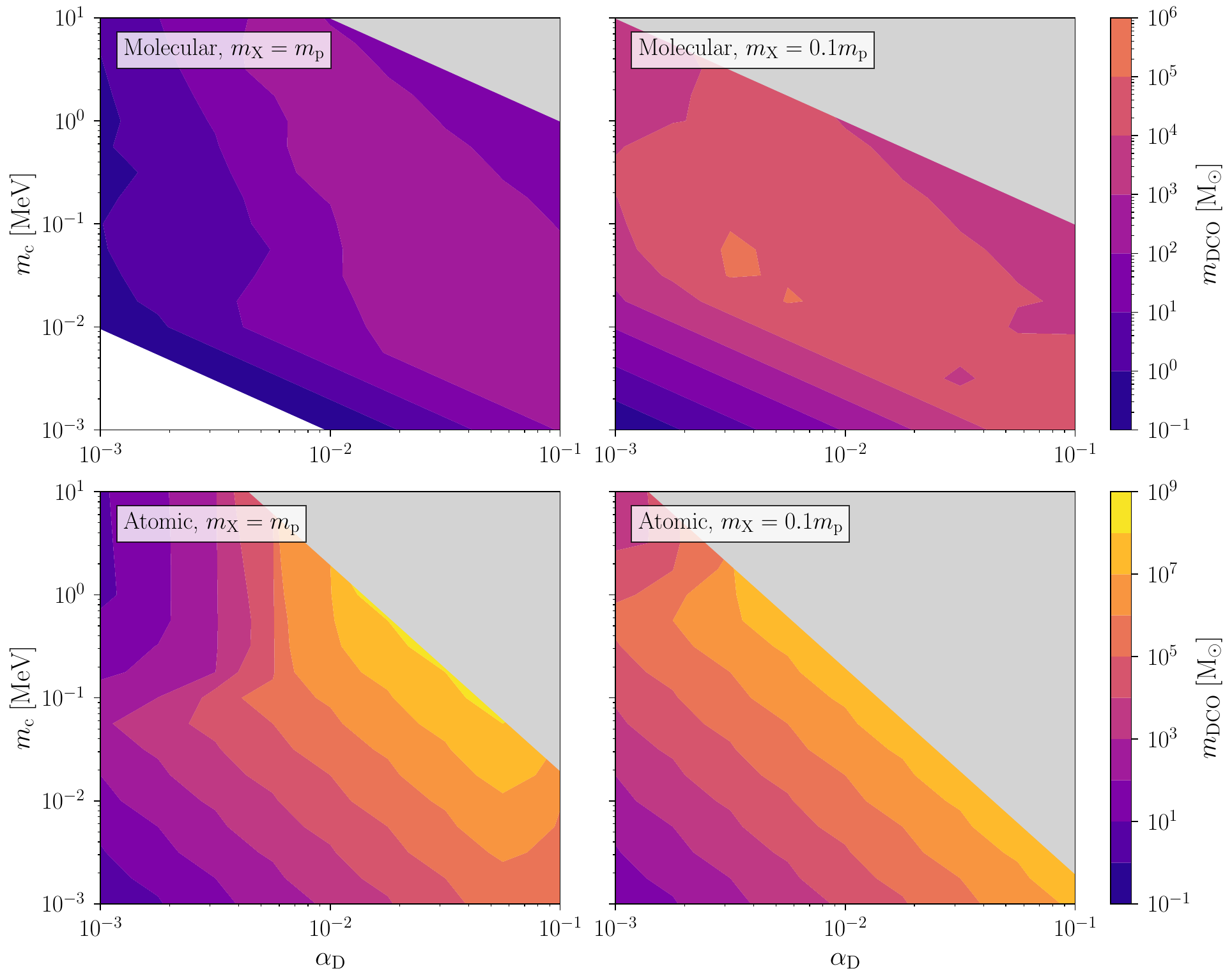}
    \caption{The \dco{} mass, $m_\dco$, as a function of $\alpha_\dark$ and $m_\del$. The top and bottom rows show masses assuming dominance of molecular and atomic cooling, respectively. The left column sets $m_\dpr = m_\pr$ and the right column sets $m_\dpr = 0.1m_\pr$. In all panels, we fix $\epsilon_\ddm=0.05$ and $\epsilon_\disk=0.01$. In the gray regions, the height of the \ddm{} disk exceeds that of the MW disk, and our procedure does not apply.}
    \label{fig:dco-mass}
\end{figure}

The star mass $m_\dco$, core temperature $T_{\core}$, and surface temperature $T_\dco$ grow as the star accretes. We describe the star by the Eddington $\beta$ model, which considers the star to have a constant ratio, $\beta$, of gas pressure to total pressure throughout its profile. This is a standard technique for SM stars~\cite{1926ics..book.....E,hansen2012stellar}, and the application to \dco{}s is justified further in \refscite{Fernandez:2022zmc,future-mass-function}.
If the star can grow indefinitely, then it will accrete all the mass around it, and the final mass will be the cloud mass given by \cref{eq:maxDDMStar}. However, three effects stop the growth of the protostar:
\begin{enumerate}
    \item 
    As the protostar grows, emissions from the protostar itself ionize the gas around it up to a radius $r_{\ion}$ known as the Str\"omgren radius.  This radius is set by competition between the recombination rate, $\alpha_{\rec}$, and the ionizing flux, $S$, which can be inferred from $T_\dco$ and $R_\dco$ via the Stefan--Boltzmann law~\cite{Ghalsasi:2017jna,McKee:2007yx,Hosokawa:2011qa}. As the ionizing luminosity increases, the Str\"omgren radius grows past the disk radius, and the gas falling onto the disk is ionized and heated,  effectively shutting down accretion onto the protostar due to the large outward pressure that forms from the temperature difference in the gas~\cite{Hosokawa:2011qa,McKee:2007yx}. Thus, the condition for accretion to terminate is given by~\cite{McKee:2007yx,Hosokawa:2011qa,future-mass-function}
    \begin{equation}
        \label{eq:rionOverrd}
        \frac{r_{\ion}}{r_\disk} \simeq
        \sqrt{
            \frac{32 \pi G m_\dco(t) m_\dpr^2 S(m_\dco(t))}
            {\alpha_{\rec} \dot m_\dco^2}
        } = 1 \, .
    \end{equation}
    We denote the time at which the above condition is satisfied as $t_\ion$. At $t_\ion$, the star stops accreting, and the final mass of the star is thus limited to $m_\dco^\ion = m_\dco^0 + \dot m_\dco t_\ion$.
    \item As the protostar grows, the central temperature becomes comparable to the dark electron mass, $T_\core(t) \sim m_\del/10$. At this point, if the mass of the star is beyond a critical mass $M_\mathrm{crit} \simeq \qty{260}{M_\odot} \times (m_\pr/m_\dpr)$, the protostar will experience core collapse through pair instability~\cite{Fernandez:2022zmc}. The protostar then collapses into a BH, suppressing further accretion.\ignorespaces
    \footnote{The BH can then further grow via Eddington-limited accretion, but we focus on parameter space with large $\alpha_\dark$ and small $m_\del$ in order to form a substantial dark disk. In this regime, Eddington-limited accretion is very slow, and further accretion can be neglected.} 
    The time at which the star experiences pair instability and collapses into a BH is given by $t_{\PI}$. The resulting \dco{} is then a BH with mass given by $m_\dco^{\PI} = m_\dco^0 + \dot m_\dco t_{\PI}$.
    \item Finally, the age of the Milky Way, $t_\milkyway \approx \qty{e10}{\year}$, limits the maximum time the star can accrete. If the star continues to evolve, it will accrete up to a mass given by $m_\dco^{\mathrm{age}} = m_\dco^0 + \dot m_\dco t_\milkyway$.
\end{enumerate}
The final mass of the star is then given by
\begin{equation}
    m_\dco = \min\Bigl\{
        m_\dco^{\ion}\,,\;
        m_\dco^{\PI}\,,\;
        m_\dco^{\mathrm{age}}\,,\;
        M_\cloud
    \Bigr\}    .
\end{equation}
We show the final masses of the dark compact objects in \cref{fig:dco-mass}. In the remainder of this work, we adopt parameters for the Milky Way and stellar disk as listed in \cref{tab:scan-parameters}.

\begin{table}
    \centering
{
\renewcommand{\arraystretch}{1.1}
\begin{tabular}{c l l @{\hspace{2cm}} c l l} 
\toprule
Quantity & Value & Refs\@. & Quantity & Value & Refs\@. \\
    \midrule
        $R_{\disk}^{\milkyway}$
           & \qty{2.15}{\kilo\parsec}
           & \cite{2022MNRAS.513.4130L,2025ApJ...990..203I}
        &
        $\height_{\disk}^{\milkyway}$
           & \qty{0.3}{\kilo\parsec}
           & \cite{2022MNRAS.513.4130L, 2025ApJ...990..203I}
    \\
        $M_{\disk}^{\milkyway}$
            & \qty{5e10}{M_\odot}
            & \cite{2025ApJ...990..203I, Cautun:2019eaf}
        &
        $M_\halo^\milkyway$
            & \qty{e12}{M_\odot}
            & \cite{Posti:2019tbp, 2022ApJ...925....1S, 2024OJAp....7E..50K}
    \\
        $R_\odot$
            & \qty{8.23}{\kilo\parsec}
            & \cite{Leung:2022dno}
        &
        $\height_\odot$
            & \qty{17}{\parsec}
            & \cite{2017MNRAS.465..472K}
    \\
        $\sigma_{\height}^2 / \sigma_R^2$
            & 0.3
            & \cite{2021MNRAS.506.1761S, 2019MNRAS.489..176M, 2022ApJ...932...28V}
    \\\bottomrule
\end{tabular}
}
\caption{The SM parameter values used in exchange rate calculations. Here $R$ and $\height$ refer to the radial and vertical coordinates of the disk, respectively, and $(R_\odot, \height_\odot)$ denote the coordinates of the Solar System.}
\label{tab:scan-parameters}
\end{table}

\section{Three-body encounters and the exchange rate}
\label{sec:three-body}
Our proposed observable is the population of star--\dco{} binaries formed by exchange of a \dco{} with one component of an existing stellar binary. In this section, in order to determine the properties of this population, we compute the exchange rate as a function of binary and \dco{} parameters. The features of three-body encounters of this kind are the subject of an enormous literature. (See \textit{e.g.}\ \refcite{2006tbp..book.....V} for an introduction, along with the seminal papers of Heggie~\cite{Heggie:1975rcz} and Hills~\cite{1975AJ.....80..809H}.) Here we follow an approach similar to that of \refcite{Bhalla:2024jbu}.

\subsection{Systematics of three-body encounters}
The formation of a star--\dco{} binary is not the only possible outcome of a close encounter. A gravitational three-body encounter involving a stellar binary system and a single \dco{} can result in one of five possible outcomes: capture, hardening, softening, disruption, or exchange, each of which is explained below. In principle, star--\dco{} binaries can also be formed by other channels besides three-body scattering. Thus, before we compute the exchange rate, we explain why three-body exchange is the most favorable process for observations.

First, note that in the absence of a nongravitational interaction between baryons and \ddm{}, forming a star--\dco{} binary directly, without first forming a star--star binary, is an extremely rare process. In the SM, stellar binaries are generally bound in their initial state: multiple stars form by fragmentation from the same gas cloud or protostar~\cite{1993ASSL..177..355B,1999ASSL..240..155T,2011psf..book.....B}. But without a dissipative channel for DM-SM interactions, baryonic gas clouds and \ddm{} clouds will not generally be mutually bound. If the star and \dco{} are not bound in the initial state, one of the two must lose energy during a close encounter in order to form a binary. This also requires a dissipative channel. Such dissipation can be provided by many-body dynamical friction, but at the typical densities of stellar environments, this is very inefficient compared to hard scatters with a third body.

Now, consider a binary system of stars labeled by 1 and 2, and an incoming \dco{} with the label 3. The role of star 1 is to carry away excess energy from the \pair23 pair to allow them to bind. \textit{Exchange} is the case in which this energy transfer is large enough that the \dco{} becomes bound to star 2, and star 1 becomes unbound from the system, leaving a star--\dco{} binary and a free outgoing star. This process has been studied as a probe of fast-moving \dco{}s by \refcite{Bhalla:2024jbu}. The other processes correspond to different regimes of energy transfer. First, the \pair12 binary can transfer energy to the \dco{}, which is the \textit{hardening} process: the binary becomes tighter as a result of the encounter. However, hardening is very inefficient when the velocity $v_3$ of the \dco{} is larger than the orbital velocity $v_{12}$ of the binary itself~\cite{Gould:1990bk,Quinlan:1996vp,Bhalla:2025xce}, which makes this observationally unfavorable. Conversely, the \dco{} can transfer energy to the binary without itself becoming bound, widening the binary. This is \textit{softening,} which has historically been used as a probe of primordial black holes and other fast-moving \dco{}s~\cite{Penarrubia:2010pa,Penarrubia:2016ltr,Ramirez:2022mys,2009MNRAS.396L..11Q,Brandt:2016aco,Tyler:2022rxi}. Significant or repeated softening leads to \textit{disruption,} in which the energy transfer is large enough to fully unbind the \pair12 binary, leaving all three components free. However, extracting constraints from softening or disruption requires an accurate model of the initial distribution of binary separations, and of the other dynamical influences that might cause them to harden, limiting the regime of applicability. The last possibility is \textit{capture,} in which the \dco{} loses enough energy to become bound to the system without unbinding the binary. One might hope to observe triple systems of two stars and one \dco{} from capture, and this has indeed been studied by \refscite{Lehmann:2020yxb,Lehmann:2022vdt}. However, such systems are unstable: the three-body capture process is time-reversible, and a subsequent close encounter eventually results in the ejection of one of the three bodies. For all intents and purposes, triple systems from capture can be regarded as long-lived intermediate states contributing to the exchange process.

As with many sectors of the three-body problem, the rate of exchanges is difficult to compute from first principles. However, the problem has been studied numerically~\cite{1990AJ.....99..979H,Heggie:1996bs}, and some general results are well understood. Exchange is most efficient when the \dco{} has a low velocity relative to the binary---which is precisely why it constitutes the ideal observable for the \ddm{} case rather than the standard CDM case---and a mass at least as large as those of the binary components.
Additionally, exchanges are most efficient when the kinetic energy of the incoming \dco{} does not exceed the binding energy of the initial binary. Otherwise, disruption can dominate: all three objects can escape to infinity. Thus, we evaluate the rate of exchanges only where the velocity of the third object is less than a critical velocity $v_{\max}$ defined such that the total energy of the three-body system is zero in the rest frame of their barycenter, given by
\begin{equation}
    \label{eq:vmax}
    v_{\max} =
    \sqrt{\frac{Gm_1m_2m_{123}}{m_{12}m_3a}} \approx
    (\qty{30}{\kilo\meter/\second})
    \left(\frac{a}{\qty{1}{\au}}\right)^{-1/2}
    \left(
        \frac{1}{\qty{1}{M_\odot}}
        \frac{m_1 m_2}{m_3}\frac{m_{123}}{m_{12}}
    \right)^{1/2}
    .
\end{equation}
Here, $m_1$ and $m_2$ denote the masses of the primary and secondary components of the initial binary system, $m_3$ refers to the mass of the \dco{}, $a$ is the initial semimajor axis of the binary, $m_{12} = m_1+m_2$, $m_{123} = m_1+m_2+m_3$, and $v_3$ is the velocity of the \dco{} in the frame of the binary barycenter. Subject to this restriction, the exchange probability $\mathcal P_\exchange$ has been computed numerically for a small range of parameters in a seminal work by Heggie, Hut, and McMillan~\cite{Heggie:1996bs}, which provides a fitting function for the cross section in the hard-binary limit:
\begin{multline}    
    \label{eq:sigma-exchange-fit}
    \Sigma_\exchange^{\mathrm{Heggie}}(v_3) \approx
    \qty{13.9}{\au^2}
    \left(\frac{a}{\qty{1}{\au}}\right)
    \left(\frac{v_3}{\qty{10}{\kilo\meter/\second}}\right)^{-2}
    \left(\frac{m_3}{m_{13}}\right)^{5/2}
    \left(\frac{m_3}{m_{12}}\right)
    \frac{(m^4_{12}m_{23}m_{123})^{1/6}}{\qty{1}{M_\odot}}  \\ 
    \exp\Bigl\{
        3.70 + 7.49\mu_1-1.89\mu_2-15.49\mu_1^2-2.93\mu_1\mu_2 - \\
        2.92\mu_2^2+3.07\mu_1^3+13.15\mu_1^2\mu_2-5.23\mu_1\mu_2^2+3.12\mu_2^3
    \Bigr\}
    ,
\end{multline}
where $\mu_1=m_1/m_{12}$ and $\mu_2=m_3/m_{123}$. Note the $v_3^{-2}$ scaling, which is crucial for sensitivity to slow-moving \dco{}s from \ddm{}.

However, since we wish to study exchanges across a wide range of masses and velocities, we must first validate this fitting function for a variety of regimes appropriate for our parameter space. We perform that validation numerically in the next subsection.

\subsection{Numerical calibration}
\label{sec:n-body}
We validate \cref{eq:sigma-exchange-fit} by numerical evaluation, largely following the procedure of \refcite{Fregeau:2004if}. Schematically, the exchange cross section can be written as
\begin{equation}
    \label{eq:sigma-exchange-probability}
    \Sigma_\exchange = \mathcal{P}_\exchange \times \pi b_{\max}^2,
\end{equation}
where $b_{\max}$ is some large impact parameter beyond which energy transfer between the binary and the \dco{} is always negligible, and $\mathcal{P}_\exchange$ represents the probability of the exchange process to occur for an impact parameter within $b_{\max}$. All of the three-body physics is embedded in $\mathcal P_\exchange$, which is dependent on all of the kinematical parameters of the three-body system. We then evaluate $\mathcal{P}_\exchange$ numerically from a suite of simulations that randomly sample the space of initial conditions.

\begin{figure}
    \centering
    \includegraphics[width=\linewidth]{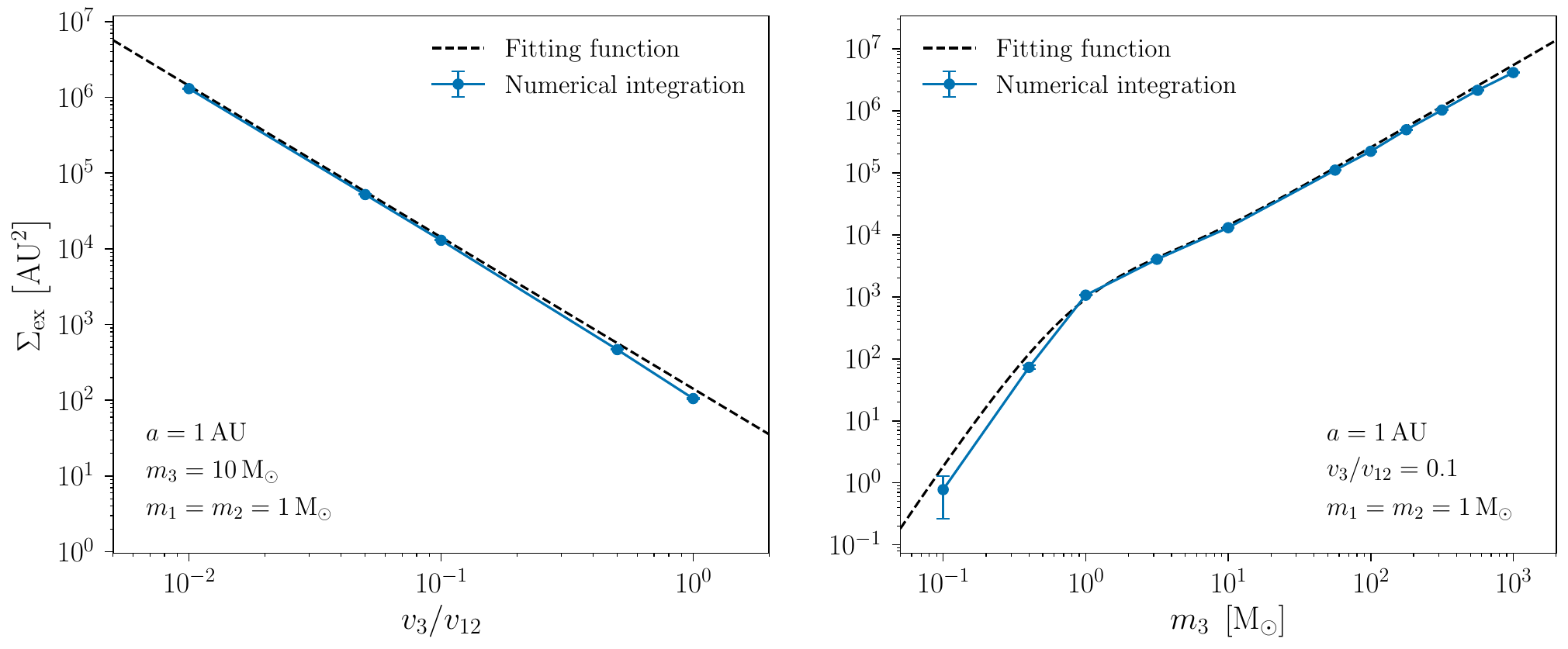}
    \caption{\textit{Left:} Exchange cross section as a function of the \dco{}'s velocity for a \dco{} of mass $m_3 = \qty{10}{M_\odot}$ with binary components of equal mass $m_1 = m_2 = \qty{1}{M_\odot}$. \textit{Right:} Exchange cross section as a function of the \dco{} mass at fixed relative velocity. The velocity of the \dco{} at infinity is fixed to $v_3 = 0.1\times v_{12}$, \textit{i.e.}, an order of magnitude smaller than the binary's orbital velocity. Our numerical results are indicated by the blue points, while the dashed black curve shows the fit of \cref{eq:sigma-exchange-calibrated}. Poisson error bars are shown for all points, but are too small to be visible in almost all cases.}
    \label{fig:cross-section-velocity-mass}
\end{figure}

\begin{figure}
    \centering
    \includegraphics[width=\linewidth]{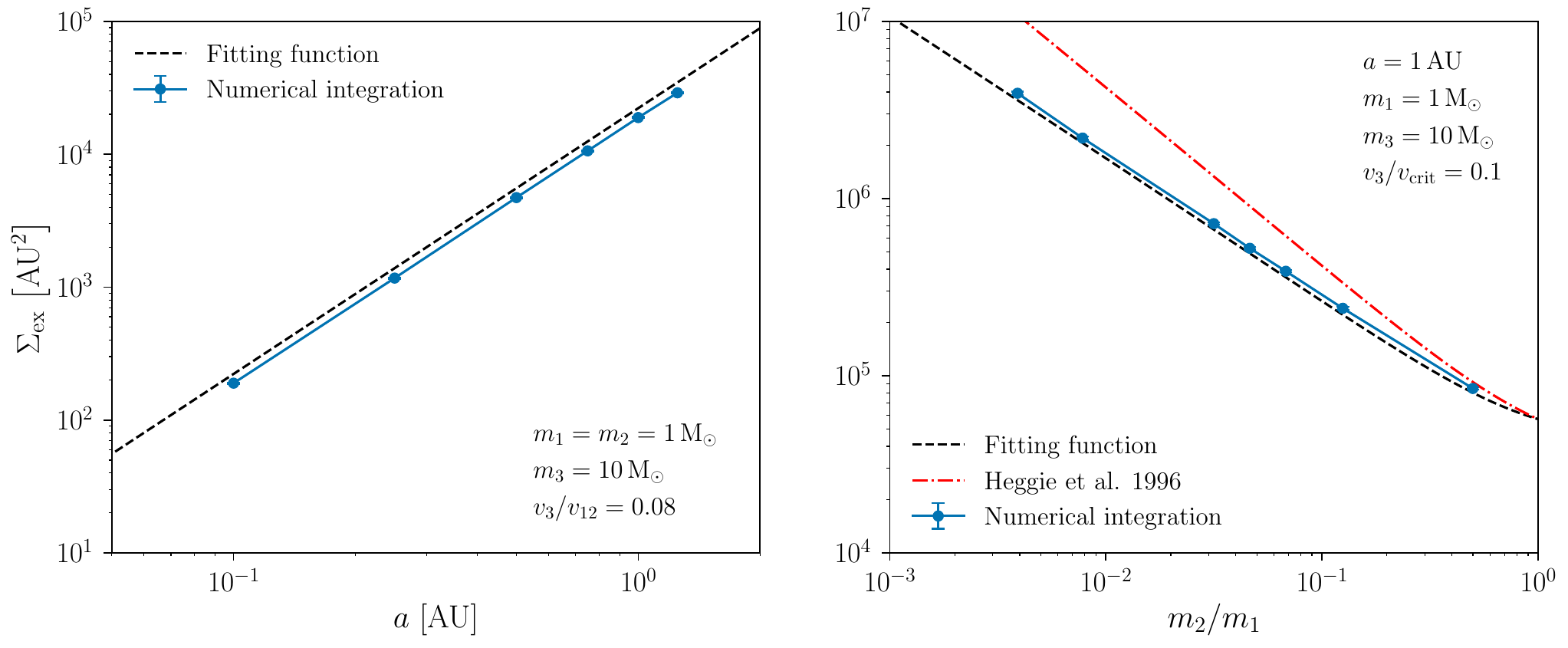}
    \caption{\textit{Left:} Exchange cross section as a function of the semimajor axis for a \dco{} of mass $m_3 = \qty{10}{M_\odot}$ with binary components of equal mass $m_1 = m_2 = \qty{1}{M_\odot}$. \textit{Right:} Exchange cross section as a function of the mass ratio of the binary components at fixed semimajor axis. The relative velocity of the \dco{} at infinity is fixed to $v_3 = 0.08\times v_{12}$. Our numerical results are indicated by the blue points, while the dashed black curve shows the fit of \cref{eq:sigma-exchange-calibrated}.
    In the right panel, the original fitting function of \refcite{Heggie:1996bs} (\cref{eq:sigma-exchange-fit}) is indicated by the dot-dashed red curve. Poisson error bars are shown for all points, and are generally too small to be visible.}
    \label{fig:CxA}
\end{figure}

We perform these simulations using the $N$-body simulation code \textsc{rebound}~\cite{Rein:2011uw}. We configure a binary system with components of mass $m_1$ and $m_2$ and semimajor axis $a$. The eccentricity is selected at random, assuming that the binaries follow a thermal distribution, with pdf $f(e)=2e$. The orbital parameters of the binary, namely, the cosine of the inclination $\cos\theta_i \in (-1,1)$; the longitude of the ascending node $\Omega \in (0,2\pi)$; the argument of pericenter $\omega \in (0,2\pi)$; and the mean anomaly $M \in (0,2\pi)$ are chosen at random for each encounter. The \dco{}, with mass $m_3$, is initially placed at $(x,y,z)=(0,0, 100a)$, and is assigned a negative radial velocity. The radial velocity $v_r$ at $(x,y,z)=(0,0, 100a)$ is set to a value such that the DCO has an asymptotic velocity $v_3$ at infinity. The transverse velocity is chosen to fix the impact parameter $b$, which is in turn randomly sampled for each simulation.

To sample $b$, we must first choose $b_{\max}$ in \cref{eq:sigma-exchange-probability}.
Due to gravitational focusing, the impact parameter is related to the pericenter distance by
\begin{equation}
    b^2=r_{\peri}^2\left(1+\frac{2Gm_{123}}{r_{\peri}v_3^2}\right)
    ,
\end{equation}
so $b_{\max}$ is related to the maximum value of $r_\peri$ in this expression. We choose $b_{\max}$ to fix $r_\peri$: for sufficiently large $r_\peri$, energy exchange between the binary and the \dco{} is suppressed, and thus the probability of exchange becomes negligibly small. Accordingly, as $r_\peri$ is increased, the computed exchange cross section first increases and then converges. For \dco{}s much more massive than either of the binary components, we choose $r_{\peri, \max} = 20 \times a$, which is sufficient for convergence at the largest masses we study in this work. For less massive \dco{}s, we find that $r_{\peri, \max} = 5 \times a$ is sufficient, and reduces the required computational time. We then sample $b^2$ uniformly on $(0,b_{\max}^2)$, reflecting the probability measure on impact parameters for random trajectories. Consequently, the exchange cross section is
\begin{equation}
    \Sigma_{\exchange} =
    \pi r_{\peri,\max}^2\left(
        1+\frac{Gm_{123}}{r_{\peri,\max}v_3^2}
    \right)\frac{N_{\exchange}}{N_{\total}}
    ,
\end{equation}
where $N_\exchange$ is the number of parameter points ending with an exchange in a suite of $N_\total$ simulations. We compute $N_{\exchange}$ from a set of \num{e4} simulated encounters between the binary and the \dco{}. The system is evolved using \textsc{rebound}'s implementation of the \textsc{ias15} integrator~\cite{2015MNRAS.446.1424R}. We stop integration when one of the three objects involved in the encounter exits a sphere of radius $100a$ with a positive energy and a positive radial velocity. The configuration of the remaining bound pair then determines whether an exchange has occurred. Encounters in which all three objects are mutually unbound are classified as disruptions and are not counted in $N_\exchange$.

Note that the exchange cross section as approximated in \cref{eq:sigma-exchange-fit} is the cross section for the event in which a single object of mass $m_1$ is ejected. To determine the total number of events that lead to the formation of a new binary system comprised of a dark compact object, it is necessary to include the cross section for the ejection of star 1 \textit{and} for the ejection of star 2. Comparisons between the resulting total cross section and our numerical simulations are shown in \cref{fig:cross-section-velocity-mass,fig:CxA}. We find that \cref{eq:sigma-exchange-fit} provides an acceptable fit to the numerical results across much of the parameter space of dark compact object mass and velocity, as well as the separation of the initial binary components. However, the fit becomes unreliable for mass ratios $q\equiv \min\{m_1,m_2\}/\max\{m_1,m_2\}$ that are not $\mathcal O(1)$. Empirically, we find an additional power-law dependence on $q$ with an index of $1/5$. We thus modify the fitting function of \cref{eq:sigma-exchange-fit} accordingly, and we compute our results using a cross section of the form
\begin{equation}    
    \label{eq:sigma-exchange-calibrated}
    \Sigma_\exchange(v_3) =
    q^{1/5} \times
    \Sigma_\exchange^{\mathrm{Heggie}}(v_3)
    .
\end{equation}

\subsection{Exchange rate in the Milky Way}
We now compute the exchange rate itself, and predict the number of exchanged objects---\textit{i.e.}, star--\dco{} binaries---expected to be present in the MW. This allows us, in turn, to estimate the number of objects expected to be observed in the \Gaia{} catalog.

In general, the star--\dco{} exchange rate per stellar binary (\textit{i.e.}, in units of inverse time) is given by
\begin{equation}
    \Gamma_\exchange =
        n_3(v_{\max})\,\Bigl\langle \Sigma_\exchange(v_3)\times v_3\Bigr\rangle
    ,
\end{equation}
where $n_3(v_{\max})$ denotes the number density of \dco{}s with $v_3 < v_{\max}$. Following the convention in the three-body scattering literature, we introduce the velocity-independent cross section $\Sigma_0\equiv\Sigma_\exchange(v_3) \times v_3^2$ to isolate the $v_3$-dependence. Then $\Gamma_\exchange = n_3(v_{\max})\Sigma_0\bigl\langle 1/v_3\bigr\rangle$, so we can write the exchange rate at fixed $m_3$ as
\begin{equation}
    \Gamma_\exchange =
    n_3(v_{\max}) \, \Sigma_0
    \int_0^{2\pi}\du\phi
    \int_0^\pi \sin\theta \dd\theta
    \int_0^{v_{\max}}v_3^2\dd v_3
        \,\frac{\mathcal{F}(v_3, \theta, \phi)}{v_3}
    ,
\end{equation}
where $\mathcal F$ is the unit-normalized 3d distribution function of $\bb v_3$, the relative velocity between \dco{}s and stellar binaries. The relative velocity is distributed as the difference between two random variables, namely the 3d velocities of the \dco{} and the binary. In general, one expects these random variables to be correlated: for example, if the stellar population forms from a baryonic cloud whose spin axis aligns with that of the DM halo, then the velocities will tend to be aligned. This would reduce the relative velocity, and substantially enhance the exchange rate. However, in order to make a conservative estimate, we assume that the velocities are uncorrelated, and treat each distribution as an independent multivariate Gaussian, as given by \cref{eq:velocity-distribution-cylindrical}. In this case, the distribution of the difference is also a Gaussian with the sum of variances in each variable, \textit{e.g.}, with $\sigma_R^2=\sigma^2_{R, \sm}+\sigma_{R, \ddm}^2$. We can then inverse-sample the $v_3$ distribution and use the Gauss--Legendre quadrature to numerically integrate over the angular degrees of freedom~\cite{Riley_Hobson_Bence_2006}. 

The exchange rate also varies with the binary parameters $m_1$, $m_2$, and $a$, so to compute the total number of exchanges in the MW, it is necessary to marginalize over these parameters. The statistics of binaries in the MW are far from settled (see \textit{e.g.}\ \refcite{2017ApJS..230...15M}), so this is at best an approximation. The total exchange rate for a particular binary is given by the sum of the exchange rates of the \dco{} with object 1 and object 2. In our definition of the exchange rate, we have assumed that the \dco{} exchanges with object 1, so the total exchange rate is the sum of the exchange rate with the binary $(m_1, m_2, a)$ and that with the binary $(m_2, m_1, a)$. We evaluate this total exchange rate for a large number of samples and average over the results to obtain the average exchange rate $\langle\Gamma_\exchange\rangle$. 

The remaining challenge is to draw a set of samples of binary parameters reflective of the binary population in the MW. We draw the mass of the heavier companion from the standard Kroupa initial mass function (IMF), with pdf given by~\cite{Kroupa:2000iv}
\begin{equation}
    \label{eq:kroupa-imf}
    p(m_1)\propto
    \begin{cases}
        m_1^{-1.3}, & \qty{0.08}{M_\odot} \leq m_1 < \qty{0.5}{M_\odot}\\
        m_1^{-2.3}, & \qty{0.5}{M_\odot} \leq m_1 \leq \qty{150}{M_\odot}.
    \end{cases}
\end{equation}
We impose a sharp cutoff above \qty{150}{M_\odot}~\cite{Kroupa:2005ph,Figer:2005gr}.
We then sample the mass ratio $q \equiv m_2/m_1 \leq 1$. For primary masses $\qty{0.8}{M_\odot} \leq m_1 \leq \qty{150}{M_\odot}$, we use the parameter distributions given by \refcite{2017ApJS..230...15M}, which provides $p(q\,|\,m_1, P)$ for binary period $P$, and $p(P\,|\,m_1)$ for binaries with $0.1 < q \leq 1$. For $\qty{0.08}{M_\odot} \leq m_1 < \qty{0.8}{M_\odot}$, we use the distributions defined in Table 1 of \refcite{2013ARA&A..51..269D} for $q$, $P$, and the binary fraction. Note that for $m_2 < \qty{0.08}{M_\odot}$, we only consider the exchange of the \dco{} with $m_2$ in our calculation of the total exchange rate, since a residual \pair23 binary would not be readily observable: object 2 would then be the only luminous component, and an object with such a low mass would be a brown dwarf and very difficult to detect. Beyond this simple prescription, we make no effort to model the \Gaia{} selection function in detail, deferring this to future work.

For $N_\star^{\sampled}$ draws, we first sample a primary mass $m_1$ from \cref{eq:kroupa-imf}. We then draw a number of companions, using the Poisson distribution of \refcite{2017ApJS..230...15M}, and count $m_1$ as a binary component if the number of companions is nonzero. For those objects which are determined to be binary components, we further sample mass ratio $q$ and separation $a$, computed from the period $P$ assuming a circular binary.  We average the exchange rate over this sample to obtain the average exchange rate \textit{per binary} per unit time, $\langle\Gamma_\exchange\rangle$. The total number of exchanges is then approximated by
\begin{equation}
    \label{eq:expected-exchanges}
    N_{\exchange}^\milkyway = N_{{\binary}}^{\milkyway}
        \times\langle\Gamma_\exchange\rangle
        \times \langle t_{\star}\rangle
    ,
\end{equation}
where $N_{{\binary}}^{\milkyway}$ is the number of binaries in the Milky Way and $\langle t_{\star}\rangle$ is the average stellar age. We conservatively take $\langle t_\star\rangle = \frac12 t_\milkyway$, \textit{i.e.}, the average stellar age assuming a uniform star formation rate, and we take $t_\milkyway=\qty{10}{\giga\year}$. We compute the number of binaries from the number of stars in the MW together with the binary fraction in our samples, \textit{i.e.}, $N_{{\binary}}^{\milkyway} = (N_{\binary}^{\sampled}/N_{\star}^{\sampled}) \times N_\star^\milkyway$ for $N_\star^\milkyway=M_\star^\milkyway / \langle m_\star \rangle$, where $M_\star^\milkyway$ is the total stellar mass of the MW and $\langle m_\star\rangle$ is the average stellar mass. We compute $\langle m_\star\rangle$ using the Kroupa IMF of \cref{eq:kroupa-imf}, and we compute $M_\star^\milkyway$ as $M_\disk^\milkyway-M^\ddm_\disk$, since the observed MW disk mass is measured gravitationally, and thus already includes any \ddm{} component. \refcite{El-Badry:2022zih} suggests that the observation of \Gaia{} BH1 implies a total population of \num{40000} star--\dco{} binaries in the MW, which would correspond under our assumptions to an average exchange rate of
\begin{equation}
    \left\langle\Gamma_\exchange\right\rangle_\mathrm{obs} =
    \qty{8e-8}{\giga\year^{-1}}
    .
\end{equation}
If \Gaia{} BH1 is interpreted as an example of an exchanged object, then $\left\langle\Gamma_\exchange\right\rangle_\mathrm{obs}$ can be regarded as the observed exchange rate of \dco{}s into stellar binaries. Alternatively, this can be treated as an estimate of the minimal exchange rate to which the currently-available \Gaia{} data is sensitive.

\section{Results}
\label{sec:results}
We now compute the exchange rates in the MW as a function of microphysical parameters, and present the number of exchanged objects expected to have formed.

\begin{figure}
    \centering
    \includegraphics[width=\linewidth]{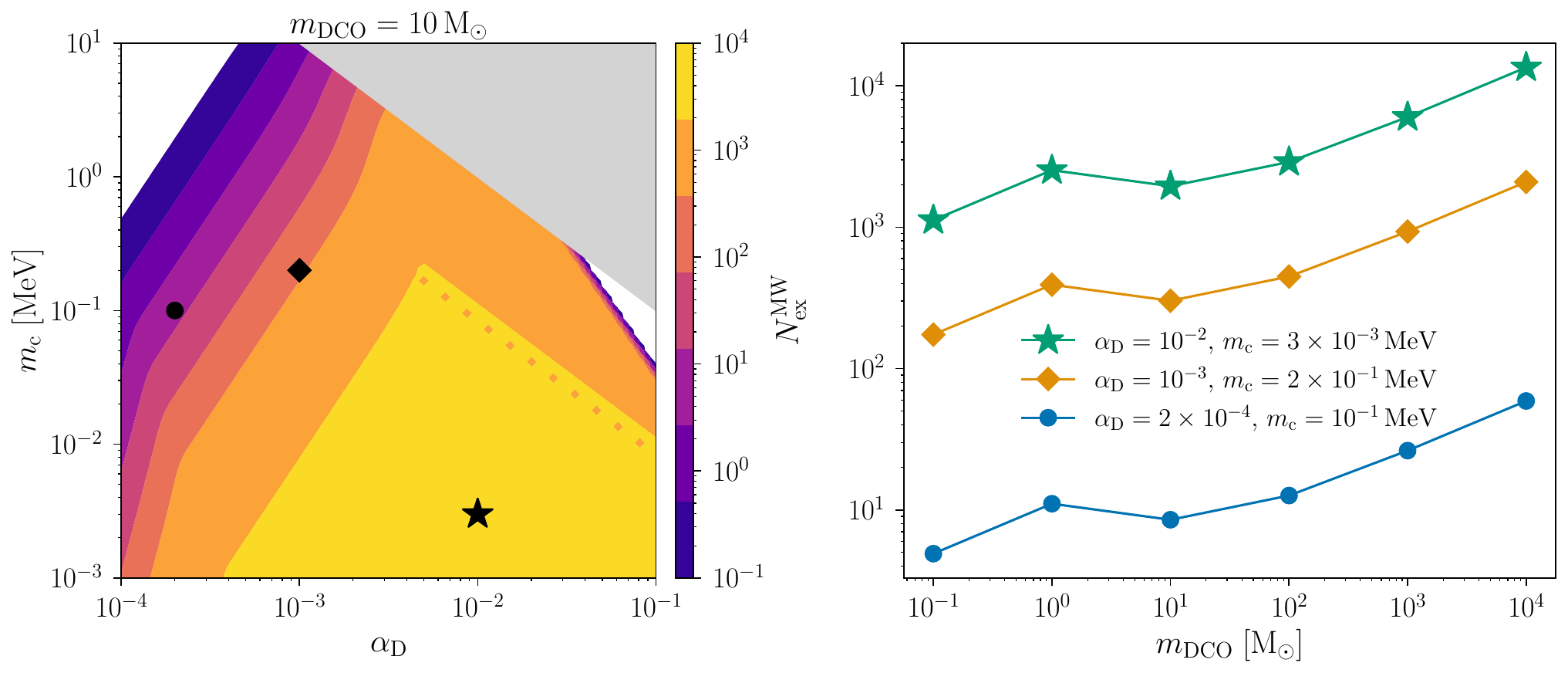}
    \caption{\textit{Left:} The number of star--\dco{} exchanges expected in the MW as a function of $m_{\del}$ and $\alpha_\dark$ at fixed \dco{} mass $m_\dco = \qty{10}{M_\odot}$. The remaining parameters are held fixed at $m_{\dpr}=0.1 m_{\pr}$, $\epsilon_\ddm=0.05$, and $\epsilon_\disk=0.01$. In the gray region, the height of the \ddm{} disk exceeds that of the MW disk, and our procedure does not apply. Benchmark parameter points corresponding to the right panel are marked with a circle, diamond, and star.
    \textit{Right:} The number of exchanges as a function of \dco{} mass for three selected parameter points with the corresponding markers in the left panel.
    }
    \label{fig:n-exchanges-mdco}
\end{figure}

First, in order to separate the role of the \ddm{} distribution from that of the \dco{} mass, we show the number of exchanges that would be expected in the MW for fixed \dco{} mass. In \cref{fig:n-exchanges-mdco}, we fix $m_\dco$ and show $N_\exchange^\milkyway$ as a function of the microphysical parameters $\alpha_\dark$ and $m_{\del}$, holding $m_{\dpr}=0.1m_{\pr}$ and $\epsilon_\ddm=0.05$ ($\epsilon_\disk=0.01$). Here, the total mass in \dco{}s is held fixed, but the exchange cross section grows somewhat faster than $m_3$, so the total number of exchanges increases mildly with the \dco{} mass (right panel). The significance of this result is that the number of exchanges is not generally dominated by the lighter objects that preferentially exchange with the low-mass tail of the stellar distribution, which would present a challenge for observation. Instead, the exchange rate prefers the formation of systems with heavy \dco{}s, which can exchange efficiently with more massive stars. The growth of the cross section stalls slightly for intermediate masses, \textit{e.g.}\ between 1 and \qty{10}{M_\odot} for the configuration shown in \cref{fig:cross-section-velocity-mass}, becoming approximately linear. Thus, the overall exchange rate plateaus in this regime, and the growth is pronounced only where the cross section becomes superlinear in $m_3$ at either very small masses $m_3 \lesssim \qty{1}{M_\odot}$ or large masses $m_3 \gtrsim \qty{10}{M_\odot}$.

The dependence of the exchange rate on the microphysical \ddm{} parameters is due to the effect of these parameters on the cooling efficiency. Larger values of $\alpha_\dark$ and smaller values of $m_\del$ produce more efficient cooling, which modifies the distribution of the \ddm{} and the resulting disk parameters. Note that at the top of \cref{fig:n-exchanges-mdco}, \textit{i.e.}, at large $m_\del$, our computation becomes unreliable: in this regime, the dark disk height $\height_\disk^\ddm$ becomes larger than the stellar thin disk height $\height_\disk^\milkyway$, so we can no longer treat the two disks as equilibrated with the same height. We thus shade this region in gray in this and subsequent figures. In the case shown here, more efficient cooling enhances the number of exchanges \textit{at fixed \dco{} mass}. Realistically, however, the \dco{} mass is not fixed, but is rather a function of the \ddm{} parameters.

\begin{figure}
    \centering
    \includegraphics[width=\linewidth]{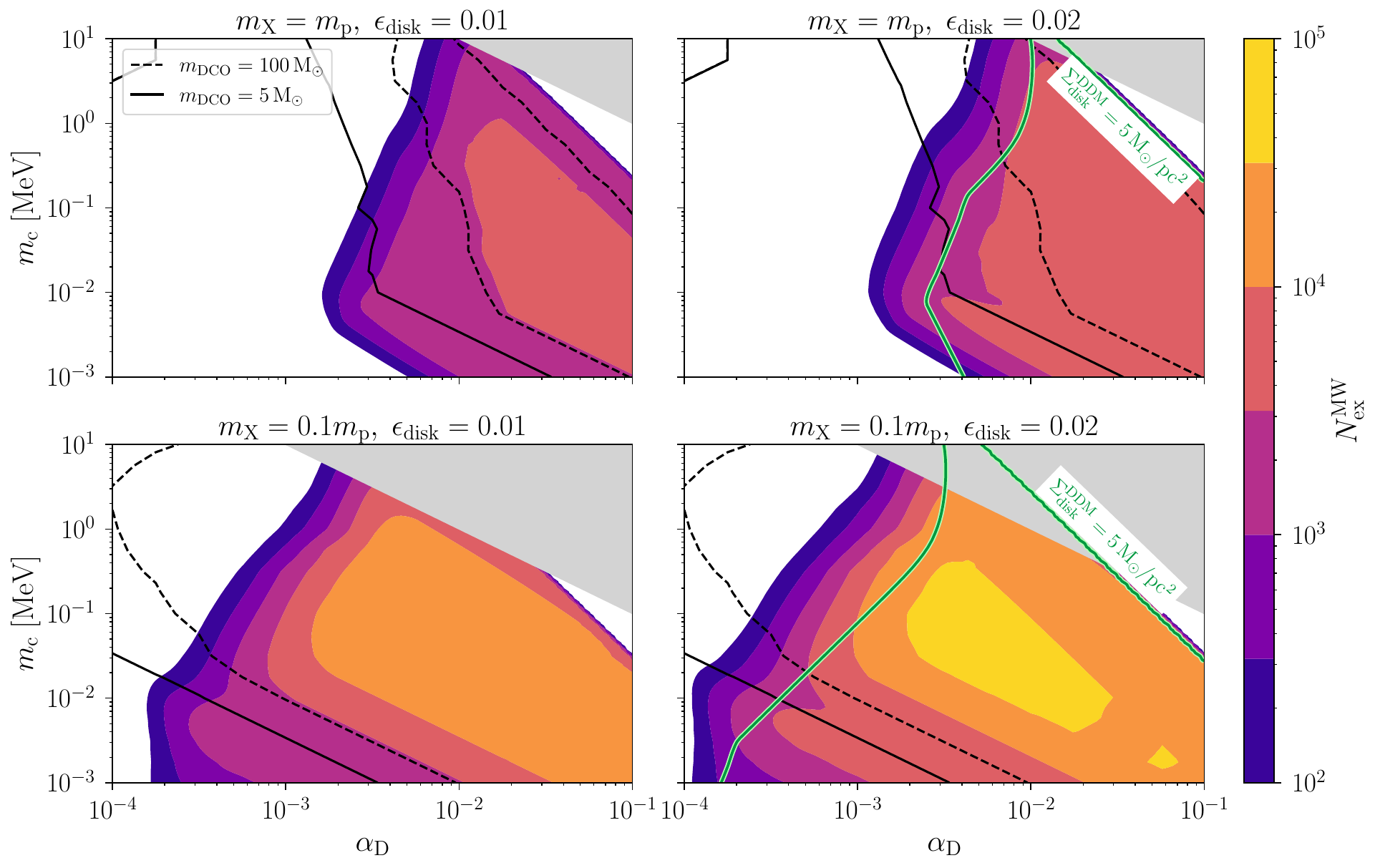}
    \caption{\textit{Binaries from molecular cooling.} The predicted number of star--\dco{} binaries in the MW as a function of $\alpha_\dark$ and $m_\del$ for four different combinations of $m_\dpr$ and $\epsilon_\disk$, assuming that molecular cooling dominates. The conversion from exchange rate to predicted number of exchanges is done using \cref{eq:expected-exchanges}. We use $\epsilon_\disk/\epsilon_\ddm=0.2$ for all panels. In the right two panels, the green curve indicates where the surface density of the disk, $\Sigma_\disk^\ddm$, reaches \qty{5}{M_\odot/\parsec^2}. This is not a direct constraint on our scenario (see text), but the region below this curve may be accessible to future dynamical constraints on dark disks. No such parameter space occurs in the left two panels, and throughout the entire parameter space, $\Sigma_\disk^\ddm$ never exceeds \qty{10}{M_\odot/\parsec^2}.
    In the gray regions, the height of the \ddm{} disk exceeds that of the MW disk, and our procedure does not apply. Black lines show the \dco{} mass under the assumption that molecular cooling is efficient.}
    \label{fig:n-exchanges-molecular}
\end{figure}

\begin{figure}
    \centering
    \includegraphics[width=\linewidth]{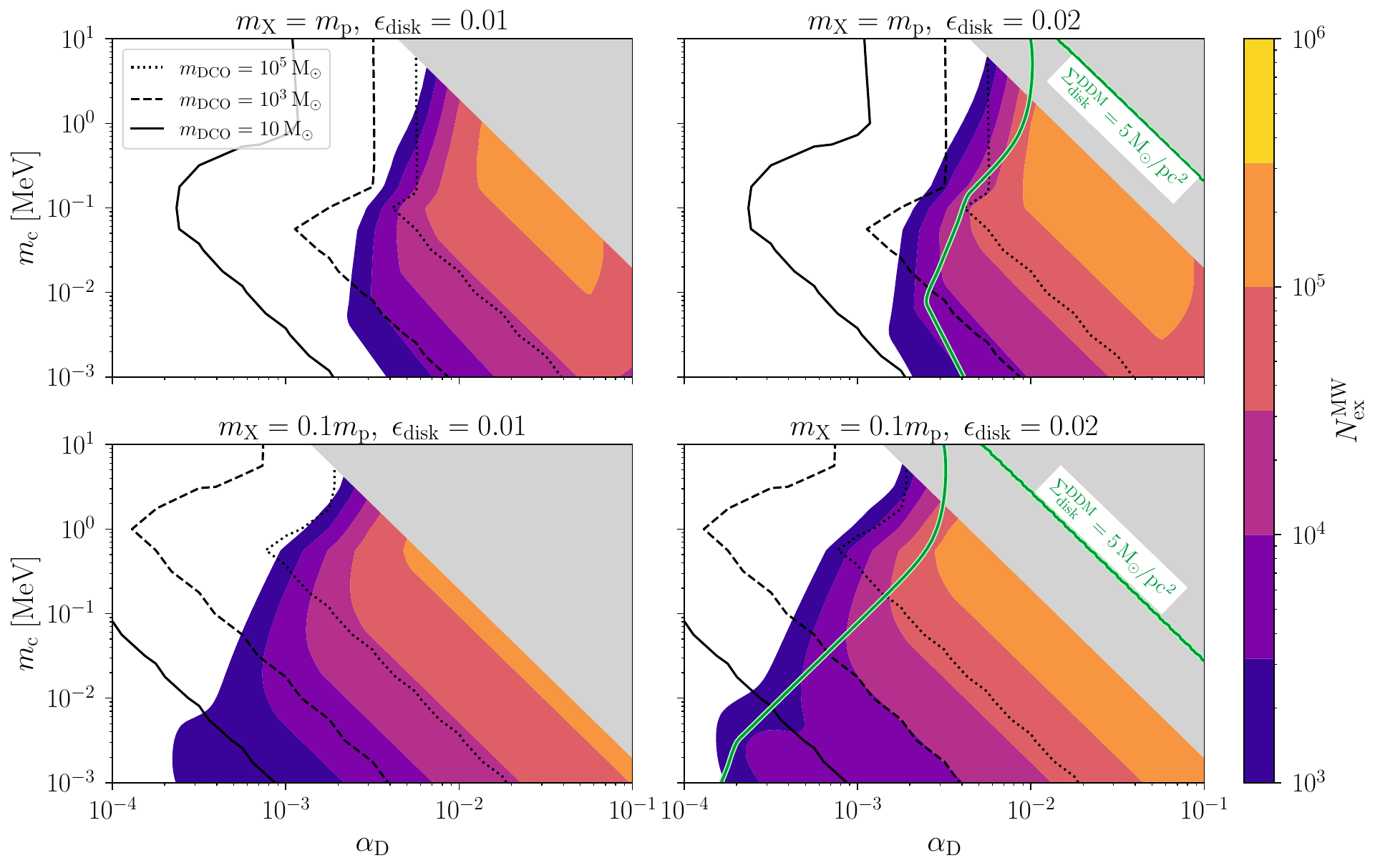}
    \caption{\textit{Binaries from atomic cooling.} The predicted number of star--\dco{} binaries in the MW as a function of $\alpha_\dark$ and $m_\del$ for four different combinations of $m_\dpr$ and $\epsilon_\disk$, as in \cref{fig:n-exchanges-molecular}, now assuming that atomic cooling dominates over molecular cooling. The conversion from exchange rate to predicted number of exchanges is done using \cref{eq:expected-exchanges}. We use $\epsilon_\disk/\epsilon_\ddm=0.2$ for all panels. In the right two panels, the green curve indicates where the surface density of the disk, $\Sigma_\disk^\ddm$, reaches \qty{5}{M_\odot/\parsec^2}. This is not a direct constraint on our scenario (see text), but the region below this curve may be accessible to future dynamical constraints on dark disks. No such parameter space occurs in the left two panels, and throughout the entire parameter space, $\Sigma_\disk^\ddm$ never exceeds \qty{10}{M_\odot/\parsec^2}.
    In the gray regions, the height of the \ddm{} disk exceeds that of the MW disk, and our procedure does not apply. Black lines show the \dco{} mass under the assumption that atomic cooling is dominant.}
    \label{fig:n-exchanges-atomic}
\end{figure}

Thus, in \cref{fig:n-exchanges-molecular,fig:n-exchanges-atomic}, we show the result of a consistent calculation that sets the \dco{} mass and distribution function from microphysical parameters. Up to the approximations made in this work, the exchange counts shown in \cref{fig:n-exchanges-molecular,fig:n-exchanges-atomic} represent our full prediction for the number of detectable exchanged objects in the MW. \Cref{fig:n-exchanges-molecular} assumes that molecular cooling dominates in setting the \dco{} mass, indicated by the black curves in the figure. \Cref{fig:n-exchanges-atomic} assumes that atomic cooling dominates, and the black curves indicate the corresponding \dco{} masses. In most of the parameter space of interest, the \dco{} mass is quite large for both cooling channels: the mass decreases below \qty{1}{M_\odot} only at the left edge of the efficient exchange region. For each of the four parameter combinations shown in each figure, there is a region with a large exchange count, plausibly large enough to produce---or even \textit{over}produce---objects like \Gaia{} BH1.

The two right panels of each of \cref{fig:n-exchanges-molecular,fig:n-exchanges-atomic} also include a green curve that shows where the surface density of the disk reaches $\Sigma_\disk^\ddm = \qty{5}{M_\odot/\parsec^2}$, which we have derived assuming the exponential disk model (\cref{eq:density}) with $R = R_{\odot}$, $\height = \height_{\odot}$, and $\height_\disk^\ddm = \qty{0.3}{\kilo\parsec}$. The value $\Sigma_\disk^\ddm = \qty{5}{M_\odot/\parsec^2}$ (\textit{i.e.}, $\rho^\ddm_\disk = \qty{0.6}{\giga\electronvolt/\centi\meter^3}$ in the Solar neighborhood) corresponds to the $95\%$ C.L. upper limit on the dark disk surface density placed by \refcite{Widmark:2021gqx}, which constrains the local DM density around \qty{300}{\parsec} in the radial direction around the Solar position for fixed disk height $\height_\disk^\ddm = \qty{0.05}{\kilo\parsec}$. Their analysis assumes a disk height that is lower than ours by a factor of six, meaning that the local density of the disk in our scenario is diluted by the same factor. Therefore, we stress that this result is not a direct constraint on our scenario, and serves only as an indication of what future dynamical constraints might probe: points below this green curve make a substantial contribution to the surface density and thus might be probed directly. Even outside this region, $N_\exchange^\milkyway$ can be large, matching the count of $\mathcal O(\num{e4})$ indicated by the detection of \Gaia{} BH1~\cite{El-Badry:2022zih}.

The exchange count in \cref{fig:n-exchanges-molecular,fig:n-exchanges-atomic} exhibits a number of features that trace to the behavior of the cooling and exchange rates. We first consider \cref{fig:n-exchanges-molecular}. Here, the exchange rate is first enhanced at large $\alpha_\dark$ and small $m_\del$, which is necessary to produce sufficient cooling for a substantial dark disk with low dispersion. However, at very large $\alpha_\dark$, the binding energy of \ddm{} atoms increases, suppressing ionization. At small $m_\del$, the equilibration timescale $t_\eq$ of \cref{eq:tdyn} increases, reducing the efficacy of cooling and suppressing overall exchange rate. In moving from $m_\dpr = m_\pr$ to $m_\dpr = 0.1m_\pr$, the exchange rate is enhanced due to increased efficiency, but this dependence is not universally monotonic: small $m_\dpr$ also implies small $T_\vir$, which reduces the efficacy of ionization. The difference between the exchange rates for molecular cooling (\cref{fig:n-exchanges-molecular}) and atomic cooling (\cref{fig:n-exchanges-atomic}) is due to the different mapping between microphysical parameters and \dco{} masses in these two scenarios. As shown in \cref{fig:dco-mass}, the atomic-cooling case produces \dco{} masses with a steeper dependence on $\alpha_\dark$, which is reflected in the steeper $\alpha_\dark$-dependence of the exchange rate in \cref{fig:n-exchanges-atomic}.

\section{Discussion and conclusions}
\label{sec:discussion}
In the foregoing sections, we have identified a new consequence of dissipation in dark sectors: the formation of abundant star--\dco{} binaries. These binaries form efficiently due to the same cooling physics that enables compact object formation: \ddm{} gas loses energy to produce \textit{slow-moving} compact objects, which are then subject to a large exchange cross section with SM stars. Such binaries are not only numerous, but plausibly observable: when the mass of the \dco{} is not much smaller than the mass of the visible partner, astrometric instruments such as \Gaia{} are capable of detecting the binary and measuring the properties of the dark partner.

Part of the motivation for studying such systems of course arises from \Gaia{} BH1 itself, since this object, and the more recently detected BH2 and BH3, provide three examples of observational detection of these systems. It is also possible that BH1 itself originates from a dissipative dark sector. We have shown that certain points in the \ddm{} parameter space naturally produce \dco{}s with exchange rates sufficient to account for the observed objects. However, we stress that these few objects do not provide any conclusive evidence to that effect. Indeed, the rates of SM formation channels, and whether they are sufficient to explain BH1, are still subject to debate in the literature. That said, \refscite{Kotko:2024oaw,Rastello:2023fkx} both estimated a formation rate of approximately \qty{2e-7}{M_\odot^{-1}} for BH1-like systems from SM channels only. Assuming a mass of $\mathcal{O}(\qty{e10}{M_\odot})$ in stars, both papers underpredict the rate approximately measured from BH1 by an order of magnitude~\cite{El-Badry:2022zih}.\ignorespaces
\footnote{We thank Kareem El-Badry for alerting us to this fact.} It is plausible that future observations, \textit{e.g.}, from \Gaia{} DR4, will provide enough data to meaningfully constrain the SM formation channels or to directly suggest a BSM origin for star--\dco{} binaries.

The other motivation for our work is to provide a new tool for constraining \ddm{} parameter space. Since star--\dco{} binaries are now observable, it is possible in principle to rule out points in parameter space by the nondetection of such objects. Such constraints can be robust in that they are independent of the SM formation channels, which then constitute a background. It is even possible that \dco{}s with masses slightly below \qty{1}{M_\odot} could be detected, and since no dark SM objects exist at such masses, this would constitute a background-free signal. In this work, we have shown that such constraints are plausible in the parameter space where the number of exchanges is large, which is not presently constrained by any direct means. We do not claim that these regions are ruled out by nondetection because such a claim requires appropriate modeling of the \Gaia{} selection function, which is a nontrivial task that we leave for future work.

Our procedure allows us to predict the properties of the \ddm{} thin disk, which has significant consequences for the exchange rate. However, a number of other physical effects may also modify the structure of the disk. Even for a purely baryonic thin disk, the effective disk parameters depend on stellar age, metallicity, and other factors~\cite{2022MNRAS.513.4130L, 2025ApJ...990..203I}. The same may be true for the dark thin disk as well. Additionally, equilibrium between the dark disk and the SM disk is a conservative and fairly strong assumption: if the two disks do not equilibrate, then in our parameter space of interest, the dark disk will form with $R_\disk^\ddm \simeq R_\disk^\milkyway$, but with $\height_\disk^\ddm \lesssim \qty{10}{\parsec} \ll \height_\disk^\sm$. This implies a smaller dispersion in the vertical direction, which significantly enhances exchange rates. At the same time, however, a disk with such a small thickness would be constrained by the analyses of \refscite{Buch:2018qdr,Widmark:2021gqx}, depending on the precise value of the Solar position $\height_\odot$.

Throughout our analysis, we have assumed $\epsilon_\disk/\epsilon_\ddm=0.2$. This originates from the simulations performed by \refcite{Roy:2023zar}. However, these simulations were performed for only two parameter points: $\alpha_\dark \in \{0.00984, 0.0182\}$ with $m_\del = \qty{0.281}{\mega\electronvolt}$ and $m_\dpr = \qty{1.3}{\giga\electronvolt}$. It is thus possible that this ratio is not universal in our parameter space of interest, and we leave further exploration to future work. Similarly, the simulations of \refcite{Roy:2023zar} predict $\height_\disk^\ddm \sim \mathcal{O}(\qty{0.1}{\kilo\parsec})$, which aligns with the baryonic values we adopt in this work, but this validation only applies to these specific parameter points.

It is possible that the parameter space of greatest interest to us, while not presently constrained, could be probed by other means. The dark-disk analyses of \refscite{Buch:2018qdr,Widmark:2021gqx} would constrain our parameter space if they were sensitive to larger values of $\height_{\disk}^{\ddm}$. As it stands, the dark disk $\height_{\disk}^{\ddm}$ we use is larger than the largest studied in either analysis, and the bounds are not directly applicable. At most, the surface density does not exceed \qty{10}{M_\odot/\parsec^2} at  $\height_{\disk}^{\ddm}=\qty{0.3}{\kilo\parsec}$ for any point in \cref{fig:n-exchanges-molecular}, twice the bound of \qty{5}{M_\odot/\parsec^2} at $\height_{\disk}^{\ddm}=\qty{0.05}{\kilo\parsec}$ reported in \refcite{Widmark:2021gqx}, and na\"ive extrapolation of their results suggests that constraints would be at least incomplete. It is also possible that PBH constraints could be adapted to constrain \dco{}s produced by \ddm{}, particularly at larger masses. This is nontrivial, however: PBH constraints in this regime rely on a combination of lensing, accretion, and dynamical effects (see \textit{e.g.}\ \refscite{Carr:2009jm,Carr:2020gox,Carr:2016drx,Carr:2020xqk,Carr:2021bzv,Carr:2026hot}). These all require assumptions regarding the distribution, compactness, and age of the \dco{}s, which do not necessarily match those of PBHs that could account for part of cosmological DM. However, extension of these limits to the \ddm{} case is possible: for example, \refcite{Winch:2020cju} shows that future LSST data will strongly constrain \dco{}s within the MW disk. We defer the detailed extension of PBH constraints for the \ddm{} case to future work.

With the release of \Gaia{} DR4, microlensing constraints may also constrain the available \ddm{} parameter space. Using Galactic disk stars, \refcite{Mondino:2023pnc} predicts sensitivity to $f_\dm<0.01$ for \dco{}s with mass $\qty{10}{M_\odot} \leq m_\dco\lesssim \qty{3e3}{M_\odot}$, significantly constraining \ddm{} models that could explain the \Gaia{} BH1 observations. Notably, \refcite{Mondino:2023pnc} assumed the dark compact objects formed a halo, not a disk. For Galactic disk stars, this could increase the sensitivity. We leave the projected constraints on the fraction of dark compact objects in a disk for future work. At the same time, however, \Gaia{} DR4 should expand the catalog of observed star--\dco{} binaries, significantly increasing the applicability of the present work.

The goal of this work is to provide a new probe of cooling in dark sectors based on its behavior at small scales. Ordinarily, predicting this small-scale behavior of DM presents computational challenges. Indeed, in the case of \ddm{}, the impacts on Galactic structure have historically been studied via expensive simulation suites. But the particular case of three-body dynamics offers a unique opportunity to study such small-scale behavior using classical dynamics in a regime that is relatively well understood. Historically, gravitational many-body dynamics has been one of the most powerful tools for studying dark sectors. Now, with \Gaia{}'s dark partners, few-body dynamics stands to become a valuable tool in its own right.

\acknowledgments
We thank Kareem El-Badry and Lisa Randall for helpful discussions. JL would also like to thank Rebecca Woody for helpful discussions.
Support for this work was provided by NASA through the NASA Hubble Fellowship grant \#HST-HF2-51610 awarded by the Space Telescope Science Institute, which is operated by the Association of Universities for Research in Astronomy, Inc., for NASA, under contract NAS5-26555.
The work of BB was supported by the Homer L. Dodge student scholarship and in part by National Science Foundation grant PHY-2412834. The work of BVL was supported in part by Simons Investigator Award \#929255. The research activities of KS are supported in part by the U.S. National Science Foundation under Grant PHY-2412671. KS acknowledges that this work was performed in part at the Aspen Center for Physics, which is supported by National Science Foundation grant PHY-2210452.  DD is supported by the James Arthur Postdoctoral Fellowship, BSF grant No. 2022287, NSF grant PHY-2514660 and the Maryland Center for Fundamental Physics, and is grateful to the Aspen Center for Physics, the Mainz Institute for Theoretical Physics, and the Munich Institute for Astro- and Particle Physics, where part of this work was carried out.

Finally, we thank Tao Xu for collaboration at earlier stages of this project. Tao was involved in this work from the beginning, and contributed to interesting ideas that we will pursue together in a future study. He was originally listed as a co-author on this work. However, due to recent policy changes, the US government funding that supports BVL no longer allows bilateral collaboration with affiliates of Chinese institutions. Tao graciously agreed to remove himself from the author list. We are grateful for his contributions, and we deeply regret this outcome. BVL, KS, and JL  wish to publicly protest and express their condemnation of this policy, which strikes at the heart of the open exchange of ideas that makes science a vehicle for peace and progress.

\bibliographystyle{JHEP}
\bibliography{references}

\end{document}